\documentclass[12pt,nofootinbib]{article}
\pdfoutput=1
\usepackage{slashed}
\usepackage{amsmath,amssymb,graphicx,multicol} 
\usepackage{epsf,color}
\usepackage[nosort]{cite}
\usepackage{cancel}
\usepackage{framed}
\usepackage{multirow}

\newcommand{\Tr}{\text{Tr}}

\newcommand{\ferm}{\chi}
\newcommand{\rep[2]}{(\mathbf{#1},\mathbf{#2})}

\def\hhref#1{\href{http://arxiv.org/abs/#1}{#1}} 

\ifx\pdfoutput\undefined
\usepackage[dvips,bookmarks=false]{hyperref}	
\else
\usepackage{hyperref}	
\fi

\newcommand{\beq}{\begin{eqnarray}}
\newcommand{\eeq}{\end{eqnarray}}

\def\bea{\begin{eqnarray}}
\def\eea{\end{eqnarray}}
\def\baa{\begin{array}}
\def\eaa{\end{array}}
\def\cO{{\cal O}}
\def\l{\left}
\def\r{\right}

\def\sl#1{\mathord{\not\mathrel{{\mathrel{#1}}}}}
\def\bit{\begin{itemize}}
\def\eit{\end{itemize}}

\newcommand{\centeron}[2]{{\setbox0=\hbox{#1}\setbox1=\hbox{#2}\ifdim

\wd1>\wd0\kern.5\wd1\kern-.5\wd0\fi \copy0

\kern-.5\wd0\kern-.5\wd1\copy1\ifdim\wd0>\wd1
                                       \kern.5\wd0\kern-.5\wd1\fi}}
\newcommand{\ltap}{\>\centeron{\raise.35ex\hbox{$<$}}
                               {\lower.65ex\hbox{$\sim$}}\>}
\newcommand{\gtap}{\>\centeron{\raise.35ex\hbox{$>$}}
                               {\lower.65ex\hbox{$\sim$}}\>}

\newcommand\ZZ{\hbox{\zfont Z\kern-.4emZ}}
\font\zfont = cmss10 

\begin{document}
\begin{titlepage}
\begin{flushright}
\end{flushright}


\begin{center}
{\Large \bf  
New Prospects for Higgs Compositeness in $h \to Z \gamma$
}
\end{center}
\vskip0.5cm

\renewcommand{\thefootnote}{\fnsymbol{footnote}}
\begin{center}
{\large Aleksandr Azatov, Roberto Contino, Andrea Di Iura, Jamison Galloway~\footnote{email:  aleksandr.azatov@roma1.infn.it, roberto.contino@roma1.infn.it, jamison.galloway@roma1.infn.it, andrea.giuseppe.DiIura@roma1.infn.it}
}
\end{center}
\renewcommand{\thefootnote}{\arabic{footnote}}

\begin{center}
{\it Dipartimento di Fisica, Universit\`a di Roma ``La Sapienza'' \\
and INFN Sezione di Roma, I-00185 Rome, Italy} \\
\vspace*{0.1cm}
\end{center}

\vglue 1.0truecm

\begin{abstract}
\noindent 
We discuss novel effects in the phenomenology of a light Higgs boson within the context of composite models.  We show that large modifications may arise in the decay of a composite  Nambu-Goldstone boson Higgs to a photon and a $Z$ boson, $h \to Z\gamma$.  These can be generated  by the exchange of massive composite states of a strong sector that breaks a left-right symmetry, 
which we show to be the sole symmetry structure responsible for  governing the size of these new effects in the absence of Goldstone-breaking interactions.   
In this paper we consider corrections to the decay $h \to Z\gamma$ obtained either by 
integrating out vectors at tree level, or by integrating out  vector-like fermions at loop level.  In each case, the pertinent operators that are generated are parametrically enhanced relative to other interactions that arise at loop level in the Standard Model such as $h \to gg$ and $h \to \gamma \gamma$.  
Thus we emphasize that the effects of interest here provide a unique possibility to probe the dynamics underlying electroweak symmetry breaking, and do not depend on any contrivance stemming from carefully chosen spectra.
The effects we discuss naturally lead to concerns of compatibility with precision electroweak measurements, and we show with relevant computations that these corrections can be kept  well under control in our general parameter space.
\end{abstract}

\end{titlepage}

\section{Introduction}
\setcounter{equation}{0}

The LHC phase of data taking at 8 TeV is over
and a large collection of experimental results on the Higgs boson has been derived.
Although data still have to be fully analyzed, a clear picture seems  to be emerging: the properties of the newly discovered particle 
closely resemble those of the Standard Model (SM) Higgs boson. 
Overall, the quantitative agreement between its measured couplings  and the SM predictions is at the $20-30\%$ level~\cite{ATLAS:2013sla,CMS:yva}. 
This strongly suggests that the new particle is indeed part of an $SU(2)_L$ doublet $H$,
and that the scale of New Physics (NP) must be somewhat higher than the electroweak scale.
From this perspective it is  important to ask which observables or processes  are most sensitive to NP effects and 
where we may be likely to see deviations from the SM pattern in the future.

It is well known that Higgs processes occurring at  loop level in the SM, such as the decays $h\to \gamma\gamma$ and $h\to Z\gamma$, and
the gluon-fusion production $gg\to h$, are particularly sensitive probes of weakly-coupled extensions of the SM.
This is typically not true, however, in theories with a light Higgs where the electroweak symmetry breaking (EWSB) dynamics is strong. If the Higgs is a composite
Nambu-Goldstone~(NG) boson of a new strongly-interacting sector,  parametrically large shifts are expected in the  tree-level couplings,
while $hgg$ and $h\gamma\gamma$ contact interactions  violate the Higgs shift symmetry and are thus suppressed.
On the other hand,  a similar symmetry suppression does not hold for a $hZ\gamma $ contact interaction.

To make this point more quantitative, contributions to the $gg$, $\gamma\gamma$ and $\gamma Z$  decay rates induced by the exchange
of new particles with mass much larger than the electroweak scale
can be conveniently parametrized by local operators. 
For a Higgs doublet, the leading NP effects are parametrized by dimension-6 operators.
A complete characterization of the Higgs effective Lagrangian at the dimension-6 level has been performed  
in previous  studies~\cite{HEL1,HEL2,Grzadkowski:2010es}; see Ref.~\cite{Contino:2013kra} for a recent review. 
In the basis of the Strongly Interacting Light Higgs (SILH) of Ref.~\cite{Giudice:2007fh}, 
the CP-conserving operators relevant for the $gg$, $\gamma\gamma$ and $Z\gamma$ rates are:~\footnote{We normalize the Wilson coefficients
according to the convention of Ref.~\cite{Contino:2013kra}.}
\begin{equation}
\label{eq:SILHOP}
\begin{split}
O_{g} & =  \frac{g_S^2}{m_W^2} H^\dagger H G_{\mu\nu}^a G^{a\, \mu\nu},  \\[0.2cm]
O_{\gamma} & =  \frac{g^{\prime\, 2}}{m_W^2} H^\dagger H B_{\mu\nu} B^{\mu\nu} , 
\end{split}
\qquad
\begin{split}
O_{HW} & = \frac{i g}{m_W^2}\, (D^\mu H)^\dagger \sigma^i (D^\nu H)W_{\mu \nu}^i \\[0.2cm]
O_{HB} & =\frac{i g^\prime}{m_W^2}\, (D^\mu H)^\dagger (D^\nu H)B_{\mu \nu} \, .
\end{split}
\end{equation}
The operators $O_g$ and $O_\gamma$ contribute  respectively to the $gg$ and $\gamma\gamma$  rates, while $Z\gamma$ gets a contribution
from both $O_\gamma$ and the linear combination $O_{HW} - O_{HB}$. The additional operators 
\begin{equation}
\label{eq:SILHOP2}
O_W = \frac{ig}{2 m_W^2} D^\nu W_{\mu\nu}^i (H^\dagger  \sigma^i  {\overleftrightarrow { D^\mu}} H) \, , \qquad
O_B = \frac{i g'}{2 m_W^2} \partial^\nu B_{\mu\nu} (H^\dagger  {\overleftrightarrow { D^\mu}} H)
\end{equation}
do not mediate $h\to Z\gamma$ for on-shell photons, but their sum contributes to the $S$ parameter~\cite{ST}.
By working in  unitary gauge and focussing on terms with one Higgs boson, the operators of Eq.~(\ref{eq:SILHOP}) are rewritten as
\begin{equation}
\label{eq:ZgammaOP}
{\cal L} = \frac{c_{gg}}{2} \, G^a_{\mu\nu} G^{a\, \mu\nu} \frac{h}{v} + \frac{c_{\gamma\gamma}}{2} \, \gamma_{\mu\nu} \gamma^{\mu\nu} \frac{h}{v} 
+ c_{Z\gamma} \, Z_{\mu\nu} \gamma^{\mu\nu} \frac{h}{v}\, ,
\end{equation}
where (by $\bar c_i$ we denote the coefficient of the operator $O_i$ of the SILH Lagrangian)~\cite{Contino:2013kra}
\begin{equation}
\begin{gathered}
c_{gg} = 8 (\alpha_s/\alpha_2) \, \bar c_g \, , \quad c_{\gamma\gamma} = 8 \sin^2\!\theta_W \, \bar c_{\gamma}\, , 
\\[0.2cm]
c_{Z\gamma} = - \tan\theta_W  \left[ (\bar c_{HW} - \bar c_{HB}) + 8 \sin^2\!\theta_W  \, \bar c_\gamma \right]\, ,
\end{gathered}
\end{equation}
and $\alpha_2 = \sqrt{2} G_F m_W^2/\pi$.
The expression of the partial decay width to $Z\gamma$, including the SM  contribution and the correction from  Eq.~(\ref{eq:ZgammaOP}) 
is given in Appendix~\ref{app:decayrate}.

Let us perform a naive estimate of the size of the effects mediated by the operators of Eq.~(\ref{eq:SILHOP}). By simple dimensional 
analysis, the Wilson coefficients $\bar c_i$ scale as $1/M^2$, where $M$ is the characteristic NP scale.
Their contribution to the  Higgs decay rates is thus suppressed by a factor $\sim (m_W^2/M^2)$, where $m_W\approx m_h$ is the energy scale
of the process. If the operators~$O_i$ are generated  by the tree-level exchange of new particles with unsuppressed non-minimal
couplings to photons and gluons, one naively expects a correction
\begin{equation}
\frac{\Delta \Gamma_{tree}}{\Gamma_{SM}} \approx \frac{m_W^2}{M^2} \, \frac{16\pi^2}{g^2}\, .
\end{equation}
Unsuppressed non-minimal couplings can arise if the new heavy states are bound states of some new strong dynamics, 
see for example the recent discussion of Ref.~\cite{Jenkins:2013fya}. 
On the other hand,  in weakly-coupled UV completions of the SM as well as in some strongly-coupled  constructions such as Holographic Higgs models,
the massive states have suppressed higher-derivative couplings.
In this case the operators $O_i$ are generated only at the cost of a loop  factor ($g_*^2/16\pi^2$)~\cite{Giudice:2007fh}, where $g_*$
denotes the coupling strength of the Higgs boson to the new states:
\begin{equation} \label{eq:dGammaloop}
\frac{\Delta \Gamma_{loop}}{\Gamma_{SM}} \approx \frac{m_W^2}{M^2} \, \frac{g_*^2}{g^2} \sim \frac{v^2}{f^2} \, .
\end{equation}
In the last identity we have defined $1/f \equiv g_*/M$.
Large corrections are thus possible in strongly-coupled theories, where 
the Higgs boson is a composite of the new dynamics and $1 \ll g_* < 4\pi$.
However, a  composite Higgs  is naturally light only if it is a NG boson of a spontaneously broken symmetry ${\cal G} \to {\cal H}$
of the strong sector. In this case  the scale $f$ must be identified with the associated decay constant,
and shifts of order $(v/f)^2$ are  expected  in the tree-level Higgs couplings to SM vector bosons and fermions 
from the Higgs non-linear $\sigma$-model Lagrangian.
At the same time, exact invariance under ${\cal G}/{\cal H}$  transformations, 
which include a Higgs shift symmetry $H^i \to H^i + \zeta^i$, 
forbids the operators $O_g$ and $O_\gamma$. For the latter to be generated the shift symmetry must be broken by some weak coupling~$g_{\not G}$,
so that the naive estimate of $\bar c_g$, $\bar c_\gamma$, and hence their contribution to the decay rates to $gg$ and $\gamma\gamma$, is further suppressed 
by an  extra factor  $(g_{\not G}^2/g_*^2)$.
Conversely, the operators $O_{HW}$ and $O_{HB}$ are  invariant under the Higgs shift symmetry, and the naive estimate (\ref{eq:dGammaloop}) 
holds for the $h\to Z\gamma$ rate. 
It follows that for a composite NG Higgs  the largest NP effects are expected to arise from shifts to the tree-level Higgs couplings
and from the contact $hZ\gamma$ interaction~\cite{Giudice:2007fh}. 
From this perspective, a precise measurement of the $Z\gamma$ rate is of crucial importance.

It is the purpose of this paper to quantitatively study  the $h\to Z\gamma$ decay rate in the context of composite Higgs models.
As implied by the  discussion above, the leading effects are captured by 
neglecting the explicit breaking of the Goldstone symmetry due to 
weak couplings of the elementary fields to the composite sector. We thus work in this limit in the following as it simplifies the calculations,
and concentrate on the contributions of  pure composite states
within minimal $SO(5)/SO(4)$ theories.
The next Section contains a brief discussion on the effective operator
basis for $SO(5)/SO(4)$ theories and the role played by the $P_{LR}$ parity for the $h\to Z\gamma$ decay.
In Section~\ref{sec:calculation} we compute the effective $hZ\gamma$ vertex generated by the tree-level exchange of spin-1 resonances and by
the 1-loop exchange of composite fermions. As a byproduct of our $hZ\gamma$ calculation we derive in Section~\ref{sec:Sparameter} the  correction to the $S$
parameter from loops of pure composite fermions.
We report our numerical results and discuss them in Section~\ref{sec:numericalanalysis}.
Useful formulas are collected in Appendices~\ref{app:decayrate}-\ref{sec:spectralfunction}, while
Appendix~\ref{sec:fccwz} contains a discussion of different formalisms commonly adopted to describe fermionic resonances. 
Finally, in Appendix~\ref{sec:masseigenstate}
we describe how the calculation of the 1-loop  contribution to $h\to Z\gamma$ from heavy fermions can be performed in full generality in the basis 
of mass eigenstates, without resorting to the approximation made in the main text.

\section{Effective Lagrangian for $SO(5)/SO(4)$ composite Higgs models and the role of $P_{LR}$}
\label{sec:operators} 
\setcounter{equation}{0}

The  effective Lagrangian for $SO(5)/SO(4)$ composite Higgs theories was discussed in Ref.~\cite{Contino:2011np}, where a complete list of four-derivative 
operators was given in the formalism of Callan, Coleman, Wess, and Zumino  (CCWZ)~\cite{CCWZ}. Here we closely follow the notation 
of Ref.~\cite{Contino:2011np}, although we adopt a different operator basis which is more transparently matched onto the SILH basis of Ref.~\cite{Giudice:2007fh}. 
At  $O(p^4)$ in the derivative expansion
there are seven independent CP-conserving  operators~\footnote{There are additionally four CP-odd  operators that can be written as
\begin{equation}
\tilde O_3^\pm = \Tr\!\left[ \Big(\tilde E^L_{\mu\nu} E^{L\, \mu\nu} \pm \tilde E^R_{\mu\nu} E^{R\, \mu\nu} \Big)^2 \right]\, , \qquad\quad
\tilde O_4^\pm = \Tr\!\left[ \big(\tilde E^L_{\mu\nu} \pm \tilde E^R_{\mu\nu}\big)\, i [d^\mu , d^\nu] \right] \, ,
\end{equation}
where $\tilde E_{\mu\nu} = \epsilon_{\mu\nu\rho\sigma} E^{\rho\sigma}$.
In particular $\tilde O_4^-$ contributes to $h\to Z\gamma$. 
Although these operators can be included straightforwardly, in this paper we will focus  on the CP-conserving ones for simplicity.
}
\begin{align}
{\cal L} & = \frac{f^2}{4} \Tr\!\left[d_\mu d^\mu \right] + \sum_i c_i O_i \\[0.3cm]
\label{eq:CCWZbasis}
\begin{split}
O_1 & = \Tr\!\left[ d_\mu d^\mu \right]^2 \\[0.1cm]
O_2 & = \Tr\!\left[ d_\mu d_\nu \right] \Tr\left[ d^\mu d^\nu \right] \\[0.1cm]
O_3^\pm & = \Tr\!\left[ (E^L_{\mu\nu})^2 \pm (E^R_{\mu\nu})^2 \right] \\[0.1cm]
O_4^\pm & = \Tr\!\left[ \left(E^L_{\mu\nu} \pm E^R_{\mu\nu}\right) i [d^\mu , d^\nu] \right] \\[0.1cm]
O_ 5 & = \sum_{a_L =1}^3 \Tr\!\left( T^{ a_L} [d_\mu, d_\nu]\right)^2 - \sum_{a_R =1}^3 \Tr\left( T^{ a_R} [d_\mu, d_\nu]\right)^2 \, , 
\end{split}
\end{align}
where 
$d_\mu(\pi) = d_\mu^{\hat a}(\pi) T^{\hat a}$, $E^{L}_\mu(\pi) = E_\mu^{a_L}(\pi) T^{a_L}$, 
$E^{R}_\mu(\pi) = E_\mu^{a_R}(\pi) T^{a_R}$ are the CCWZ covariant  functions---transforming respectively as an adjoint and gauge fields of $SO(4)$---of the NG field $\pi(x) = \pi^{\hat a}(x) T^{\hat a}$.  Explicitly,
\beq
\label{eq:dEdef}
-i U^\dagger(\pi) D_\mu U(\pi) = d_\mu + E_\mu^L + E_\mu^R \, ,
\eeq
where $U(\pi) = \exp (i \sqrt{2}\,  \pi (x)/f)$.
The field strength $E_{\mu \nu}^{L,R}$ is constructed from a commutator of covariant derivatives $\nabla_\mu = \partial_\mu + i (E_\mu^{L}+ E_\mu^R)$; 
see Ref.~\cite{Contino:2011np}  for further details. 
Here $T^{a_L}$, $T^{a_R}$ are the generators of the unbroken $SO(4) \sim SU(2)_R \times SU(2)_R$, while those of $SO(5)/SO(4)$ are denoted as $T^{\hat a}$.
The SM electroweak vector bosons weakly gauge
an  $SU(2)_L \times U(1)_Y$ subgroup of $SO(5) \times U(1)_X$, where the $U(1)_X$ does not participate in the dynamical breaking, but  is needed to correctly reproduce
the hypercharges of the SM fermions. It is convenient to define 
the tree-level vacuum such that  the electroweak group is fully contained in the unbroken 
$SO(4) \times U(1)_X$, with $Y = T^{3R} + T^X$. The true vacuum will in general be misaligned with this direction by an angle $\theta$ due to the radiatively induced 
potential of the NG bosons.

The operators in Eq.~(\ref{eq:CCWZbasis}) have been conveniently defined to be even or odd under a  parity $P_{LR}$ which exchanges the $SU(2)_L$ and $SU(2)_R$ 
comprising  the unbroken $SO(4)$. Under $P_{LR}$ the NG bosons transform as $\pi^{\hat a}(x) \to -\eta^{\hat a} \pi^{\hat a}(x)$, 
with $\eta^{\hat a} = \{ 1,1,1,-1\}$, which implies $d_\mu^{\hat a} \to -\eta^{\hat a} d_\mu^{\hat a}$, $E^L_{\mu} \leftrightarrow E^R_\mu$~\cite{Contino:2011np}.
Ordinary parity is thus the product $P = P_0 \cdot P_{LR}$, where $P_0\,$$: (t, \vec x) \to (t,-\vec x)$ is the usual spatial inversion. 
Under $P_{LR}$, the operators $O_{1,2}$ and $O^+_{3,4}$  are even, whereas $O^-_{3,4}$ and $O_5$  are odd.
Expanding in the number of NG fields
it is easy to match the operators of Eq.~(\ref{eq:CCWZbasis}) with the dimension-6 operators of the SILH Lagrangian by noticing that
\begin{equation}
\label{eq:expansion}
\begin{split}
d_\mu & \sim D_\mu H + \dots \\
E_{\mu\nu} & \sim A_{\mu\nu} + \frac{1}{f^2} \left[ D_\mu \!\left( H^\dagger  i {\overleftrightarrow { D_\nu}} H \right)  - (\mu \leftrightarrow \nu) \right]+ \dots 
\end{split}
\end{equation}
where the gauge fields entering into the field strength $A_{\mu\nu}$ and the covariant derivative $D_\mu$ are those of $SU(2)_L \times U(1)_Y$.
From Eq.~(\ref{eq:expansion}) one can see that $O_3^\pm$ and $O_4^\pm$  correspond respectively to $O_W\pm O_B$ and $O_{HW} \pm O_{HB}$. 
The leading terms in  the expansion of $O_1$, $O_2$, and $O_5$ 
are instead of dimension~8, meaning that these operators do not have a 
counterpart in the SILH Lagrangian of Ref.~\cite{Giudice:2007fh}. The exact relations and the connection between our basis and that of 
Ref.~\cite{Contino:2011np}  are reported  in  Appendix~\ref{app:opbases}. 
Notice that there is no operator in Eq.~(\ref{eq:CCWZbasis}) corresponding to $O_\gamma$ and $O_g$, since the latter explicitly
break the $SO(5)$ global symmetry.
It then follows that the only (CP-conserving) operator which gives a $hZ\gamma$ contact coupling is~$O_4^-$:
\begin{equation}
\label{eq:cZga}
c_{Z\gamma} = g^2 \sin^2\!\theta \, c_4^-  \, .
\end{equation}
Notice also that only $O_3^+$ contributes to the $S$ parameter:
\begin{equation}
S = - 32\pi \sin^2\!\theta \, c_3^+ \, .
\end{equation}

It is not an accident that the  $hZ\gamma$ coupling follows from a $P_{LR}$-odd operator. Since the abelian $U(1)_X$ subgroup 
factorizes with respect to the non-linearly realized  $SO(5)$, the photon and $Z$ fields enter into the operators of Eq.~(\ref{eq:CCWZbasis}) 
only through the weak gauging of the $U(1)_L \times U(1)_R$ subgroup. By formally assigning the transformation rules $W_\mu^{3} \leftrightarrow B_\mu$,
$g \leftrightarrow g^\prime$,  the $P_{LR}$ symmetry is exact even after turning on the neutral gauge fields.
By the above rules, the $Z$ field is odd while the photon and the Higgs boson are even under $P_{LR}$, so that the decay $h\to Z\gamma$ can be
mediated only by an odd operator.  By the same argument, $S$ is an even quantity under $LR$ exchange (it is proportional to the coefficients
of the unitary-gauge operator $W_{\mu\nu}^3 B^{\mu\nu}$), and it is consistently induced by the $P_{LR}$-even operator $O_3^+$.

It is interesting to notice that in the limit of unbroken $SO(5)$ symmetry the RG running of $c_4^-$, as well as that of the coefficient of any 
$O(p^4)$ odd operator, vanishes due to the $P_{LR}$ parity.
While the effective operators are generated at some high-energy scale $M$ by the exchange of massive states, the Wilson coefficients
appearing in the expressions of low-energy observables, like in Eq.~(\ref{eq:cZga}) for the $Z\gamma$ decay rate, must be evaluated at the 
typical scale of the process, $\mu \ll M$.
The running of the Wilson coefficients from $M$ down to $\mu$ originates from 1-loop logarithmically divergent diagrams
constructed with $O(p^2)$ vertices,
\begin{equation}
c_i(\mu) = c_i(M) + \frac{b_i}{16\pi^2} \log\frac{M}{\mu}\, ,
\end{equation}
where $b_i$ are $O(1)$ numbers. Since however $P_{LR}$ is an accidental symmetry of the $O(p^2)$ Lagrangian~\cite{Contino:2011np}, it follows that there cannot be 
any running   from loops of NG bosons in the case of $P_{LR}$-odd operators, i.e. $b_i = 0$.~\footnote{From the viewpoint of the SILH Lagrangian of 
Ref.~\cite{Giudice:2007fh}, this argument shows that there cannot be any contribution to the RG evolution of $O_{HW}-O_{HB}$ and $O_{W}-O_{B}$ from $O_H$.
In fact, the authors of Ref.~\cite{Elias-Miro:2013gya} showed that even the combination $O_{HW}+O_{HB}$ is not renormalized by $O_H$.}
An RG evolution is in general induced by loops of transverse vector bosons,~\footnote{For the calculation of the RG running relevant to 
$h\to \gamma\gamma$ and $h\to Z\gamma$ see Refs.~\cite{Grojean:2013kd,Elias-Miro:2013gya}.} since the weak gauging explicitly breaks $P_{LR}$ 
at the $O(p^2)$ level. 
This is however a subleading electroweak  effect which we neglect for simplicity in this paper.
One naively expects $c_i(M) \sim 1/16\pi^2$ for  operators  generated at the scale $M$ with an extra loop suppression, as in the case
of $O_4^-$ for a minimally-coupled UV theory. 
Hence, if the RG running was non-vanishing, the leading contribution to the  Wilson coefficient could   come from long-distance 
(log-enhanced)  effects rather than from high-energy threshold corrections. 
For $P_{LR}$-even operators generated at  tree-level,  like $O_3^+$ for example, one instead estimates  $c_i(M) \sim 1/g_*^2$, so that the threshold contribution
dominates over the RG evolution  as long as $g_* \lesssim 4\pi/\sqrt{\log(M/\mu)}$.

If  $P_{LR}$  is an exact  invariance of the strong dynamics, then  $c_4^-(M)$ vanishes for unbroken~$SO(5)$. 
In this case the only source of $P_{LR}$ breaking
stems from the couplings of the elementary gauge and fermion fields to the strong sector, and the $hZ\gamma$ contact interaction 
will be suppressed by a  factor $(g_{\not G}/g_*)^2$, as in the case of $h\gamma\gamma$ and $hgg$.
We will thus focus on the case in which the strong dynamics explicitly breaks the $P_{LR}$ symmetry, so that $O_4^-$ is generated at the scale $M$
even in the limit of unbroken $SO(5)$. A  $LR$-violating strong dynamics generically leads to dangerously large corrections to the $Zb\bar b$ 
vertex~\cite{Agashe:2006at}, but there are special cases where the  $LR$ breaking is  communicated to the $Zb\bar b$ coupling  in a suppressed way.
For example, it has been pointed out by the authors of Ref.~\cite{Mrazek:2011iu} that if the fermionic resonances form (only) fundamental representations of 
$SO(5)$, as in the MCHM5~\cite{Contino:2006qr}, then 
$P_{LR}$ is an accidental invariance of the lowest derivative fermionic operators relevant for $Zb\bar b$
(small $P_{LR}$-breaking effects suppressed by $m_b/m_t$ are present but can be neglected). In this case the spin-1 sector of resonances can be 
maximally $LR$ violating, and thus generate an unsuppressed  $hZ\gamma$  interaction, without leading to excessively  large shifts in the $Zb\bar b$ coupling. 
Another possibility is that the sector of fermionic resonances maximally breaks $P_{LR}$ but the shift of $Zb\bar b$  is suppressed by a small coupling.
A minimal realization of this case can be obtained for example  if the spectrum of fermionic resonances contains both fundamental and antisymmetric
representations of $SO(5)$.
In the following Section we will provide two explicit models realizing these  possibilities and compute the  contribution of the resonances to the $h\to Z\gamma$
decay rate.

\section{$h\to Z\gamma$ from pure composite states}
\label{sec:calculation}
\setcounter{equation}{0}

We calculate the contribution of pure composite states to $h\to Z\gamma$ 
by focusing on the  lightest modes and describing their dynamics by means of a low-energy effective theory. 
For our description to be valid we assume that these states are lighter
than the cutoff scale $\Lambda$ where other resonances occur, and that the derivative expansion of the effective theory is controlled by $\partial/\Lambda$.
Notice however that whenever it arises at the 1-loop level, the contribution of the lightest modes to $h\to Z\gamma$ is parametrically of the same order as that of the
cutoff states. These latter are heavier but are also expected to  be more strongly coupled than the lighter modes, so that both effects are naively of order $(v/f)^2$,
as shown by Eq.~(\ref{eq:dGammaloop}).  In this case our calculation should be considered as a more quantitative estimate of the contribution of the strong dynamics 
rather than a precise  prediction of a model.
For a more detailed discussion about the validity of this effective description we refer the reader to Ref.~\cite{Contino:2011np}, whose approach we follow in 
this paper (see also Ref.~\cite{Contino:2006nn}). 

In the fermionic sector we assume the existence of linear elementary-composite couplings, 
which leads to partial compositeness~\cite{Kaplan:1991dc,Contino:2006nn}. 
We are however interested in the effects of pure composite states, hence in the following
we work at lowest order in the elementary couplings and set them to zero.
Before EWSB the composite states fill multiplets of the linearly-realized subgroup  $SO(4) \times U(1)_X\sim SU(2)_L \times SU(2)_R \times U(1)_X$.
We will consider spin-1 resonances in the $(\mathbf{3}, \mathbf{1})_0$ and $(\mathbf{1},\mathbf{3})_0$ representations (denoted respectively $\rho^L$
and $\rho^R$), and fermionic resonances transforming as $(\mathbf{1},\mathbf{1})$, $(\mathbf{2},\mathbf{2})$, $(\mathbf{1},\mathbf{3})$ and
$(\mathbf{3},\mathbf{1})$ representations with arbitrary
$U(1)_X$ assignments.

In the following parts of this section we first derive the $hZ\gamma$ contact coupling
by integrating out the composite states  and matching to the low-energy  theory.
We will then illustrate two minimal models where maximal $P_{LR}$ breaking can occur without generating a large modification  of the $Zb\bar b$ coupling.

\subsection{Tree-level exchange of spin-1 resonances}

We begin by considering the contribution from the tree-level exchange of spin-1 composites.~\footnote{See also Ref.~\cite{Cai:2013kpa} for
a calculation of the 1-loop contribution to $h\to Z\gamma$ from vector resonances.}
We follow the vector formalism where the $\rho$ transforms non-homogeneously under $SO(5)$ transformations.
Neglecting  CP-odd operators for simplicity, the effective Lagrangian for 
$\rho_\mu^L = \rho^{a_L}_\mu T^{a_L}$ and $\rho_\mu^R = \rho^{a_R}_\mu T^{a_R}$ can be written 
as follows~(see Ref.~\cite{Contino:2011np} for more details):~\footnote{At the level of leading terms in the derivative expansion
there are  four additional CP-odd operators:
\begin{equation}
\tilde Q_{1r}  =   \epsilon^{\mu\nu\alpha\beta} \, \Tr\!\left( \rho^r_{\mu\nu} \, i [d_\alpha, d_\beta] \right)\, ,  \qquad
\tilde Q_{2r}  =  \epsilon^{\mu\nu\alpha\beta} \, \Tr\!\left( \rho^r_{\mu\nu} E^r_{\alpha\beta} \right) \, , \qquad r = L,R\, .
\end{equation}
For simplicity we will concentrate on CP-even operators in the following, although the inclusion of the CP-odd ones is straightforward. 
Notice also that we use a slightly different basis of  operators $Q_i$ compared to Ref.~\cite{Contino:2011np}, 
so as to  match more easily with the low-energy Lagrangian~(\ref{eq:CCWZbasis}).
 }
\begin{equation}
\label{eq:Lrho}
\begin{split}
{\cal L} = & -\frac{1}{4g^2_{\rho_L}} \,\Tr\!\left( \rho_{\mu\nu}^{L}  \rho^{L\, \mu\nu}\right) + \frac{m_{\rho_L}^2}{2 g_{\rho_L}^2} \,\Tr\!\left( \rho^{L}_\mu - E_\mu^{L} \right)^2
               + \alpha_{1L}\, Q_{1L} + \alpha_{2L} \, Q_{2L}\\[0.2cm]
               & -\frac{1}{4g^2_{\rho_R}} \,\Tr\!\left( \rho_{\mu\nu}^{R}  \rho^{R\, \mu\nu}\right) + \frac{m_{\rho_R}^2}{2 g_{\rho_R}^2} \,\Tr\!\left( \rho^{R}_\mu - E_\mu^{R} \right)^2
               + \alpha_{1R}\, Q_{1R}+ \alpha_{2R} \, Q_{2R}\, ,
\end{split}
\end{equation}
where we have neglected subleading terms in the derivative expansion and have defined
\begin{equation}
Q_{1r} = \Tr\!\left( \rho^r_{\mu\nu} \, i [d^\mu, d^\nu] \right)\, , \qquad
Q_{2r} = \Tr\!\left( \rho^{r\, \mu\nu} E^r_{\mu\nu} \right)\, , \qquad r = L,R\, .
\end{equation}
It is straightforward  to integrate out the spin-1 resonances at tree-level by using the equations of motion: $\rho_\mu = E_\mu + O(p^3)$.
One obtains the low-energy Lagrangian (\ref{eq:CCWZbasis}) with 
\begin{equation}
c_3^\pm = \frac{1}{2} \left[ \left( \alpha_{2L} - \frac{1}{4g_{\rho_L}^2}\right) \pm \left( \alpha_{2R} - \frac{1}{4g_{\rho_R}^2}\right) \right]\, , \qquad
c_4^\pm = \frac{1}{2} \left(\alpha_{1L} \pm \alpha_{1R} \right)\, .
\end{equation}

The $S$  parameter receives a correction both from the $\rho$ mass terms and from 
$Q_{2L}$, $Q_{2R}$~\cite{Contino:2011np}.~\footnote{Ref.~\cite{Orgogozo:2012ct} pointed out that  $Q_{2L}$, $Q_{2R}$
modify the high-energy dependence of the current-current vacuum polarizations at tree-level and can be made consistent with the UV behavior of the OPE 
only if an additional contribution to the operators $O_3^\pm$ exist, $\Delta c_3^{\pm} = - (\alpha_{2L}^2 g_{\rho_L}^2 \pm \alpha_{2R}^2 g_{\rho_R}^2)$. 
The expression of the $S$ parameter thus reads~\cite{Orgogozo:2012ct}:
\begin{equation}
\label{eq:Sparameter}
S = 4\pi \sin^2\!\theta \left[ \left(\frac{1}{g_{\rho_L}} - 2 g_{\rho_L}  \alpha_{2L} \right)^2
  + \left(\frac{1}{g_{\rho_R}} - 2 g_{\rho_R}  \alpha_{2R} \right)^2 
\right]\, .
\end{equation}
No similar issue arises with $Q_{1L}$, $Q_{1R}$.}
The vertex $hZ\gamma$, 
on the other hand, 
follows from the operators $Q_{1L}$, $Q_{1R}$  due to the $\rho$-photon mixing induced by the 
$\rho$ mass term:
\begin{equation}
\label{eq:cZgarho}
c_{Z\gamma} = \frac{g^2}{2} \sin^2\!\theta \left(\alpha_{1L} - \alpha_{1R} \right)\, .
\end{equation}
After rotating to the basis of mass eigenstates, $\rho_{\mu\nu}$ gives a photon field strength $\gamma_{\mu\nu}$, 
while  $d_\mu d_\nu$ gives $Z_\mu \partial_\nu h$.~\footnote{It is possible to diagonalize the mixing of the $\rho$ with the elementary gauge fields
by making the field redefinition $\bar\rho_\mu = \rho_\mu - E_\mu$, where $\bar\rho$ transforms as a simple adjoint of $SO(4)$. 
In this mass eigenstate basis the tree-level exchange of $\bar\rho$ does not generate a $hZ\gamma$ vertex, due to the simple fact that no appropriate 
 Feynman diagrams can be constructed. 
Instead, the $hZ\gamma$ interaction arises directly from the contribution to 
$O_4^-$ that follows from $Q_{1r}$ after replacing $\rho_{\mu\nu} = \nabla_{[\mu} \,\bar\rho_{\nu]}  + E_{\mu\nu} + i [\bar\rho_\mu , \bar\rho_\nu]$. 
This is analogous to what happens for the $S$ parameter and in fact for any observable at  leading order in the derivative expansion.
Indeed, it is easy to check that integrating out the $\bar\rho$ by means of the 
equations of motion  generates only $O(p^6)$ operators, i.e. operators with more than four derivatives.
This shows that the Lagrangian written in terms of  $\bar\rho$ must be properly supplemented with additional four-derivative terms, among which is $O_4^-$, 
in order to match the original one~\cite{Ecker:1989yg}.}
In this sense the operators $Q_{1r}$, unlike $Q_{2r}$, 
give non-minimal  couplings of the photon  to neutral particles.

The size of the correction to the $h\to Z\gamma$ decay rate depends on the value of the parameters $\alpha_{1r}$.
By assuming Partial UV completion~\cite{Contino:2011np}, so that the strength of the interactions
mediated by $Q_{1r}$ becomes of order $g_* \equiv \Lambda/f$ at the cutoff scale $\Lambda$, one estimates $\alpha_{1r} \lesssim 1/(g_\rho g_*) < 1/g_\rho^2$.
In a minimally-coupled theory, on the other hand, the operators $Q_{1r}$ carry a further loop suppression from which the more conservative estimate 
$\alpha_{1r} \sim 1/(16\pi^2)$, and in turn Eq.~(\ref{eq:dGammaloop}), follow.

\subsection{Loops of fermionic resonances}
\label{sec:fermloops}

\begin{figure}[t]
\begin{center}
\includegraphics[width=0.3\linewidth]{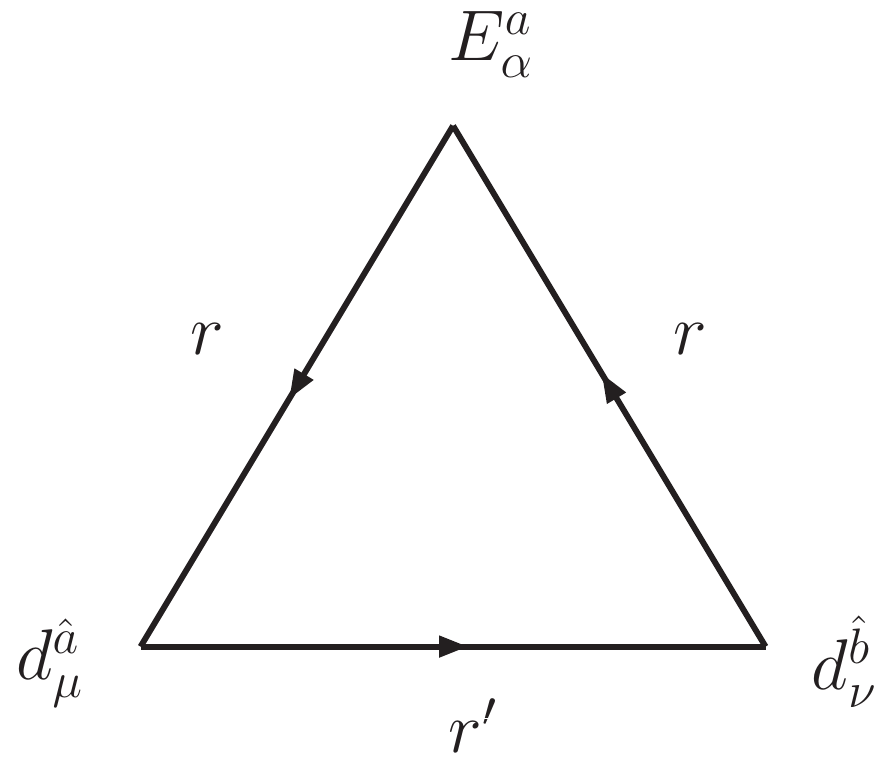}
\caption{\small 
One-loop contribution to the Green function $\langle E^a_\alpha d^{\hat a}_\mu d^{\hat b}_\nu \rangle$
from composite fermions in the representations $r$ and $r^\prime$ of $SO(4)$.
}
\label{fig:triangle}
\end{center}
\end{figure}

Composite fermions can generate the  vertex $hZ\gamma$ at 1-loop level. 
Let us consider for example the case of fermions transforming as $(\mathbf{1},\mathbf{1})$, $(\mathbf{2},\mathbf{2})$, $(\mathbf{1},\mathbf{3})$ and
$(\mathbf{3},\mathbf{1})$ under $SO(4) \sim SU(2)_L \times SU(2)_R$ 
and with arbitrary common 
$U(1)_X$ charge. 
At leading order in the derivative expansion, the  Lagrangian reads
\begin{equation}
\label{eq:Lfermion}
\begin{split}
{\cal L} = \sum_r \,\bar\ferm_r \!\left( i \slashed \nabla - m_r \right) \ferm_r
- \big[  & \zeta_{11} \, \bar\ferm_{\rep[2]{2}} \slashed d\, \ferm_{\rep[1]{1}}  + \zeta_{13} \, \bar\ferm_{\rep[2]{2}} \slashed d\, \ferm_{\rep[1]{3}}  \\
&  +\zeta_{31} \, \bar\ferm_{\rep[2]{2}} \slashed d\, \ferm_{\rep[3]{1}} + h.c. \big]\, ,
\end{split}
\end{equation}
where $r$ runs over all $SO(4)$ representations, $\zeta_{11}$, $\zeta_{13}$, $\zeta_{31}$ are $O(1)$ complex coefficients, and 
$\nabla_\mu = \partial_\mu + i (E_\mu^L + E_\mu^R)$ is the covariant derivative on $SO(5)/SO(4)$.
By integrating out the fermions and matching with the low-energy Lagrangian~(\ref{eq:CCWZbasis}), the contribution to $c_4^\pm$ comes from the one-loop
diagram of Fig.~\ref{fig:triangle} plus its crossing, where one has to sum over all possible representations~$r$,~$r^\prime$.
%

For a given diagram with fermions in the representations $r$ and $r^\prime$ of $SO(4)$, the Feynman amplitude can be expressed as 
\begin{equation}
\label{eq:AmplitudeForm}
M_{\alpha\mu\nu}^{a\hat a\hat b} = N_\chi \,  \omega^{a\hat a \hat b}_{[r, r^\prime]}\, |\zeta_{[r,r^\prime]}|^2\, I_{\alpha\mu\nu}(p_1, p_2 ; m_r, m_{r^\prime})\,  ,
\end{equation}
where $p_1$ and $p_2$ are the momenta of  $d_\mu^{\hat a}$ and $d_\nu^{\hat b}$ respectively (defined to be flowing into the corresponding vertices),  
the index $a$ runs over the adjoint of $SU(2)_L \times SU(2)_R$, and $N_\chi$ is the fermion multiplicity.
For example, for three families of colored fermions (heavy quarks) one has $N_\chi = N_c N_F =9$ (with $N_c = 3$, $N_F=3$), while $N_\chi = 12$
if there are three additional  families of colorless fermions (i.e. heavy leptons).
Here $\zeta_{[r,r^\prime]} = \zeta_{[r^\prime,r]}^*$ denotes the coupling strength of $d_\mu$
with fermions in the representations $r$ and $r^\prime$: $\zeta_{[\rep[2]{2},\rep[1]{1}]} = \zeta_{11}$,
$\zeta_{[\rep[2]{2},\rep[1]{3}]} = \zeta_{13}$, and $\zeta_{[\rep[2]{2},\rep[3]{1}]} = \zeta_{31}$, as in Eq.~(\ref{eq:Lfermion}). 
By $SO(4)$ covariance, the second factor of Eq.~(\ref{eq:AmplitudeForm}) is proportional to the $SO(4)$ generator $t^a_{\hat a\hat b}$, 
\begin{equation}
\label{eq:deflLR}
\omega^{a_L\hat a \hat b}_{[r,r^\prime]} = l^L_{[r,r^\prime]}\,  t^{a_L}_{\hat a\hat b}\, , \qquad 
\omega^{a_R\hat a \hat b}_{[r,r^\prime]} = l^R_{[r,r^\prime]}\,  t^{a_R}_{\hat a\hat b}\, ,
\end{equation}
where the coefficients $l^{L,R}_{[r,r^\prime]}$ are reported in Table~\ref{tab:groupfactors} for the fermion representations under study.
%
\begin{table}
\begin{center}
\renewcommand{\arraystretch}{1.2}
\begin{tabular}{lccccc}
$[r,r^\prime]$  \hspace{0.2cm} 
& $[\rep[2]{2},\rep[1]{1}]$ &  $[\rep[2]{2},\rep[1]{3}]$ & $[\rep[1]{3},\rep[2]{2}]$ & $[\rep[2]{2},\rep[3]{1}]$ & $[\rep[3]{1},\rep[2]{2}]$  \\ \hline 
$l^{L}_{[r,r^\prime]}$ & $1$ &  $+3/4$ & 0 & $-1/4$  & $1$ \\[0.2cm]
$l^{R}_{[r,r^\prime]}$ & $1$ &  $-1/4$ & $1$ & $+3/4$ & 0 
\end{tabular}
\end{center}
\caption{\small 
Value of the coefficients $l^{L,R}_{[r,r^\prime]}$ defined in Eq.~(\ref{eq:deflLR}) for diagrams with fermions in the $\rep[1]{1}$, $\rep[2]{2}$, $\rep[1]{3}$ and 
$\rep[3]{1}$ of $SO(4)\sim SU(2)_L \times SU(2)_R$. The coefficients not shown in the Table are vanishing.
} 
\label{tab:groupfactors}
\end{table}
%
The contribution to $c_4^\pm$ can be  extracted by expanding the loop function $I_{\alpha\mu\nu}$ at first order in the external momenta.
It is easy to show 
that $I_{\alpha\mu\nu}$ is antisymmetric under the exchange $\{\mu, p_1\} \leftrightarrow \{ \nu, p_2 \}$, so  
there are three possible Lorentz structures at linear order in the external momenta:
\begin{equation}
\label{eq:Idec}
\begin{split}
I_{\alpha\mu\nu}  = \, & A(m_r, m_{r^\prime}) \, \eta_{\mu\nu} (p_1 - p_2)_\alpha + B(m_r, m_{r^\prime}) \left( p_{1\,\mu} \eta_{\nu\alpha} - p_{2\,\nu} \eta_{\mu\alpha} \right) 
\\[0.1cm]
& + C(m_r, m_{r^\prime}) \left[ \eta_{\nu\alpha} (p_1 +p_2)_\mu - \eta_{\mu\alpha} (p_1 +p_2)_\nu \right] + O(p^3) \, .
\end{split}
\end{equation}
The functions $A$, $B$, $C$ are logarithmically divergent and their expression is given in Appendix~\ref{app:loopfuncts}.
The terms proportional to $A$ and $B$ renormalize respectively the operators $O_{d1} = \Tr[(\nabla_\mu d_\nu)^2]$ and $O_{d2} =\Tr[(\nabla_\mu d^\mu)^2]$,
which contain terms with zero, one and two $E_\mu$'s from the covariant  derivative.  
In fact, the same function $I_{\alpha\mu\nu}$ accounts for the one-loop contribution to the three-point Green function 
$\langle J^a_\alpha d^{\hat a}_\mu d^{\hat b}_\nu \rangle$,
where $J_\mu^a$ is the $SO(4)$ conserved current (see Eq.~(\ref{eq:currents})).
It is thus subject to the Ward identity
\begin{equation}
\label{eq:WI}
i (p_1 + p_2)^\alpha  I_{\alpha\mu\nu} = G_{\mu\nu}(p_1^2) - G_{\mu\nu}(p_2^2)\, ,
\end{equation}
where $G_{\mu\nu}(p^2)$ is the  $d_\mu$ self-energy:
\begin{equation}
\langle d^{\hat a}_\mu d^{\hat b}_\nu \rangle |_{amp} \equiv \delta^{\hat a\hat b} G_{\mu\nu}(p^2)  = 
\delta^{\hat a\hat b}  \eta_{\mu\nu} \, \Pi_0(p^2)  + \delta^{\hat a\hat b} p_\mu p_\nu \, \Pi_1(p^2) \, .
\end{equation}
From Eq.~(\ref{eq:WI})  it follows that
\begin{equation}
A(m_r,m_{r'}) = \Pi_0^\prime(0)\, , \qquad
B(m_r,m_{r'}) = \Pi_1(0) \, .
\end{equation}
We have checked these identities by explicitly computing the self-energy $G_{\mu\nu}$.
The coefficients of the operators $O_{d1}$, $O_{d2}$ are given by
\begin{equation}
\label{eq:cd12}
\begin{split}
c_{d1} & = \frac{N_\chi }{4} \sum_{r,r^\prime} A(m_r,m_{r^\prime}) \left( l_{[r,r^\prime]}^L + l_{[r,r^\prime]}^R \right)\, |\zeta_{[r,r^\prime]}|^2 \, , \\[0.15cm]
c_{d2} & = \frac{N_\chi}{4}  \sum_{r,r^\prime} B(m_r,m_{r^\prime}) \left( l_{[r,r^\prime]}^L + l_{[r,r^\prime]}^R \right)\, |\zeta_{[r,r^\prime]}|^2 \, ,
\end{split}
\end{equation}
where the sums are over all possible 
fermion representations $r, r^\prime$ contributing to the 1-loop diagram of 
Fig.~\ref{fig:triangle}.~\footnote{Notice that the expressions of $c_{d1}$ and $c_{d2}$ are $LR$ symmetric, as required since
the operators $O_{d1}$, $O_{d2}$ are even under $P_{LR}$. The corresponding $LR$-odd combinations vanish because the functions $A(m_r,m_{r^\prime})$,
$B(m_r,m_{r^\prime})$ are symmetric under the exchange $r\leftrightarrow r^\prime$ 
and due to the sum rule (\ref{eq:sumrule}).}
Neither of the operators $O_{d1}$, $O_{d2}$ contributes to $h\to Z\gamma$:
$O_{d2}$ can be redefined away in terms of higher-derivative operators by using the equations of motion $\nabla_\mu d^\mu = 0$, while $O_{d1}$ 
can be rewritten as
\begin{equation}
\label{eq:opidentity}
\Tr[(\nabla_\mu d_\nu)^2] \equiv O_{d1} = \frac{1}{2} \Tr[ F_{\mu\nu}^2 ] - \frac{1}{2} O_3^+ + \frac{1}{4} \left( O_2 - O_1 \right)\, ,
\end{equation}
and thus contributes to the $S$ parameter. We will discuss this further in the next section, where we perform a detailed calculation of $S$.

Finally,  the term proportional to $C$ in Eq.~(\ref{eq:Idec}) renormalizes the operator $\Tr[E_{\mu\nu} d^\mu d^\nu]$ and thus contributes to $c_4^\pm$.
We find:
\begin{equation}
\label{eq:final1loop}
c_4^\pm = \frac{N_\chi}{4}  \sum_{r,r^\prime} C(m_r,m_{r^\prime}) \left( l_{[r,r^\prime]}^L \pm l_{[r,r^\prime]}^R \right)\, |\zeta_{[r,r^\prime]}|^2 \, .
\end{equation}
Using the coefficients of Table~\ref{tab:groupfactors} and Eqs.~(\ref{eq:functionC}) 
and (\ref{eq:cZga}),  one can  derive the fermionic contribution to the 
$hZ\gamma$ vertex.
In particular, in a theory with composite fermions only in the $\rep[1]{1}$ and $\rep[2]{2}$ representations, the  contribution to~$c_4^-$ (hence to $c_{Z\gamma}$)
 vanishes identically, since $l^L_{[\rep[2]{2},\rep[1]{1}]} = l^R_{[\rep[2]{2},\rep[1]{1}]}$ and $l^{L,R}_{[\rep[1]{1},\rep[2]{2}]} =0$. 
This is expected, since the fermionic sector in this case possesses  an accidental $P_{LR}$
invariance. When fermions in the $\rep[1]{3}$ and $\rep[3]{1}$ are present, however, the contribution to $c_4^-$ is non-vanishing provided $P_{LR}$ is broken
either by the couplings ($\zeta_{13} \not = \zeta_{31}$) or in the spectrum ($m_{\rep[1]{3}} \not = m_{\rep[3]{1}}$).

It is interesting to notice that although the function $C$ is logarithmically divergent, the contribution to $c_4^-$ from Eq.~(\ref{eq:final1loop}) is finite, since the
coefficients $l_{[r,r^\prime]}^{L,R}$ satisfy the sum rule
\begin{equation}
\label{eq:sumrule}
\left( l_{[r,r^\prime]}^L - l_{[r,r^\prime]}^R \right) + \left( l_{[r^\prime,r]}^L - l_{[r^\prime,r]}^R \right) = 0\hspace{1.0cm}
\text{for any } r,r^\prime\, . 
\end{equation}
This identity can be directly checked on the coefficients of Table~\ref{tab:groupfactors}, and a simple argument shows that it holds in general 
for any pair $(r,r^\prime)$.
The proof goes as follows. 
When computing the 1-loop diagram of Fig.~\ref{fig:triangle}, it is useful to treat $E_\mu$ and $d_\mu$ as external backgrounds coupled to the fermions.
Let us then turn on $E_\mu$ along the diagonal $U(1)_{L+R}$ subgroup of $SO(4) \sim SU(2)_L \times SU(2)_R$, under which
$d_\mu^1 \pm i d_\mu^2$ has charge $\pm 1$  and $d_\mu^3$ and $d_\mu^4$ have charge~$0$.
By charge conservation there  are only two possible diagrams (plus their crossings)  as in Fig.~\ref{fig:triangle}:  one with $d^1_\mu$ and $d^2_\nu$ at the two 
lower vertices, the other with $d^3_\mu$ and $d^4_\nu$.
Since $d_\mu^{1,2,3}$ are odd while $d_\mu^4$ is even under $P_{LR}$,  the second diagram contributes to~$O_4^-$, while the first
renormalizes~$O_4^+$.~\footnote{A more direct way to see this is the following: the $SO(4)$ generators satisfy $t^{3_L}_{34} = - t^{3_R}_{34}$,
$t^{3_L}_{12} =  t^{3_R}_{12}$, which implies 
$O_4^- =-i ( t^{a_L}_{\hat a \hat b} E_{\mu\nu}^{a_L} - t^{a_R}_{\hat a \hat b} E_{\mu\nu}^{a_R}) d_\mu^{\hat a} d_\nu^{\hat b} 
\supset v_{\mu\nu} d_\mu^3 d_\nu^4$,
$O_4^+ = -i( t^{a_L}_{\hat a \hat b} E_{\mu\nu}^{a_L} + t^{a_R}_{\hat a \hat b} E_{\mu\nu}^{a_R}) d_\mu^{\hat a} d_\nu^{\hat b} 
\supset v_{\mu\nu} d_\mu^1 d_\nu^2$ in the background $v_\mu = E_\mu^{3L} = E_\mu^{3R}$.
In terms of physical fields, $O_4^-$ contains a term $\gamma^{\mu\nu} (\partial_\mu h) Z_\nu$, while  $O_4^+$ contains 
$\gamma^{\mu\nu} W^+_\mu W^-_\nu$.
}
We thus concentrate on the diagram with $d^3_\mu d^4_\nu$ and notice that the fermions circulating in the loop must
all have the same $U(1)_{L+R}$ charge. Let $\lambda_{ij}^3$ and $\lambda_{ij}^4$  be the coupling strengths of two same-charge fermions $i$ and $j$ 
respectively to $d_\mu^3$ and $d_\mu^4$ (in the fermions' mass eigenbasis). 
For a given diagram with fermions $i$ and $j$ in the loop, the log-divergent part is thus proportional to $(\lambda^3_{ij} \lambda^4_{ji})$.
Due to  the antisymmetry of the loop function,  $I_{\alpha\mu\nu}(p_1,p_2,m_i,m_j) = - I_{\alpha\nu\mu}(p_2,p_1,m_i,m_j)$, 
the log-divergent part of the crossed diagram is instead proportional to $- (\lambda^4_{ij} \lambda^3_{ji})$.
The sum then vanishes after summing over all fermions $i$, $j$ with the same charge.
This proves that there is no log-divergent  contribution to $c_4^-$ from 1-loop fermion diagrams, 
hence the sum rule (\ref{eq:sumrule}) must hold for any pair of $SO(4)$ representations $(r,r^\prime)$. 
In general, a logarithmic divergence $\log (\Lambda/m)$ is associated with the running of a Wilson coefficient from the cutoff scale $\Lambda$ down to the
fermion mass scale~$m$. The above argument thus shows that there is no RG running of $c_4^-$ induced by 1-loop diagrams of fermions
above their mass scale, while such running is present in general for $c_4^+$.

\begin{figure}[t]
\begin{center}
\includegraphics[width=0.3\linewidth]{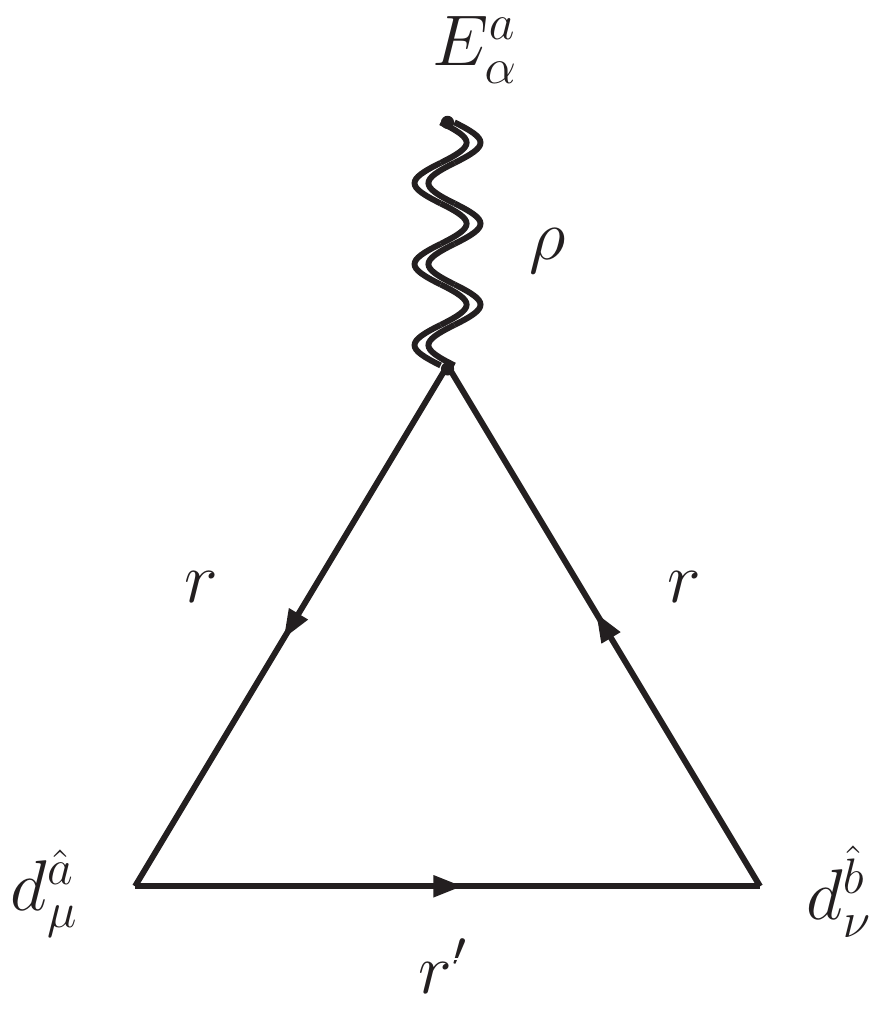}
\caption{\small 
Mixed rho-fermion contribution to the Green function $\langle E^a_\alpha d^{\hat a}_\mu d^{\hat b}_\nu \rangle$ which arises at the one-loop level.
}
\label{fig:trianglerho}
\end{center}
\end{figure}

So far we have considered 1-loop diagrams with only composite fermions. There are also diagrams where both fermions and spin-1 resonances
can circulate. At the 1-loop level, however,   $\rho^{L,R}$  can only appear external to the loop due to the mixing with $E_\mu$ from its mass term and
a coupling to the fermions of the form
\begin{equation}
\label{eq:fermrhoint}
\bar\ferm_r \left( \rho_\mu - E_\mu \right) \gamma^\mu \ferm_r \, .
\end{equation}
The corresponding diagram is shown in Fig.~\ref{fig:trianglerho}. 
%
%
%
It is easy to see that its contribution to $c_4^-$ vanishes at leading order:  integrating out the $\rho$ through the equations of motion generates
only four-fermion operators, which in turn do not contribute at the 1-loop level.
In general, the tree-level  exchange of the $\rho$ in the diagram of Fig.~\ref{fig:trianglerho} leads to a form-factor correction to the  vertex
of $E_\mu$ with the fermionic current. For $q^2 = (p_1+p_2)^2 \ll m_\rho^2$ such a form factor correction is of order $q^2/m_\rho^2$ and is thus suppressed compared
to the direct interaction from the fermions' kinetic terms.

\subsection{Two models}
\label{sec:models}

We have already alluded  to the fact that a generic $P_{LR}$-violating strong dynamics can lead to unacceptably large corrections to the $Z\bar b b$ 
vertex~\cite{Agashe:2006at}.
Here we  sketch two simple models where the  breaking of $P_{LR}$ is communicated to the $Z\bar b b$ vertex in a suppressed way, such that a sizable
correction to $h\to Z\gamma$ is phenomenologically allowed.

\paragraph{Model 1} In the first model, which is a low-energy simplified version of the MCHM5~\cite{Contino:2006qr},  the composite fermions fill  two fundamental representations 
of $SO(5)$,   with $U(1)_X$ charge $+2/3$ and $-1/3$ respectively: 
\begin{equation}
\begin{split}
\ferm_5 & = \rep[1]{1}_{2/3} + \rep[2]{2}_{2/3} \\[0.1cm]
\ferm^\prime_5 & = \rep[1]{1}_{-1/3} + \rep[2]{2}_{-1/3} \, .
\end{split}
\end{equation}
The spectrum of composite states also includes a $\rho^L$ and $\rho^R$, while we omit for 
simplicity spin-1 states transforming as bifundamentals
of $SU(2)_L \times SU(2)_R$.
The Lagrangian can be written as ${\cal L} = {\cal L}_{elem} + {\cal L}_{comp} + {\cal L}_{mix}$, where ${\cal L}_{elem}$
describes the elementary fields in isolation and the expression of the composite Lagrangian ${\cal L}_{comp}$  is as in Eqs.~(\ref{eq:Lrho}) and (\ref{eq:Lfermion}).
The term $ {\cal L}_{mix}$ accounts for the mixing of the elementary to composite fermions:
\begin{equation}
{\cal L}_{mix} =
\lambda_q \, \bar q_L P_q U(\pi) \ferm_5 + \lambda^\prime_q \, \bar q_L P_q U(\pi) \ferm_5^\prime
+ \lambda_t \, \bar t_R P_t U(\pi) \ferm_5 + \lambda_b \, \bar b_R P_b U(\pi) \ferm_5^\prime + h.c. \, ,
\end{equation}
where  $P_{q,t,b}$ project out the components of the composite fields with the  
electroweak quantum numbers of the
corresponding elementary fields.
The $P_{LR}$ invariance is taken to be maximally violated in the spin-1 sector, but is accidentally preserved  in the fermion sector. If $\lambda_q^\prime \ll \lambda_q$, 
then the $Z\bar b b$ coupling is protected from large corrections 
since for  $\lambda_q^\prime =0$ there is no operator  at leading order in the derivative expansion which can modify it~\cite{Mrazek:2011iu}.
A small $\lambda_q^\prime/\lambda_q$ can in fact naturally arise from the RG running of the full theory and explain the hierarchy 
between the top and bottom masses if $\lambda_t \simeq \lambda_b$~\cite{Contino:2006qr}.
We note in passing that at tree level the correction to $Z\bar b b$ from a $\rho^{L,R}$ is always vanishing at leading order in the derivative
expansion, as  can be easily checked by using the equations of motion $\rho_\mu  = E_\mu + O(p^3)$ in Eq.~(\ref{eq:fermrhoint}). This is because the shift 
induced by  the exchange of the $\rho$ is exactly compensated by the additional interaction 
$\bar\ferm \gamma^\mu E_\mu \ferm = \bar\ferm \gamma^\mu  (H^\dagger  i {\overleftrightarrow { D_\mu}} H + \dots)  \ferm$ required by $SO(5)$ invariance
and included in the term of Eq.~(\ref{eq:fermrhoint}).  A non-vanishing $Z\bar bb$ will however arise in general at the 1-loop level in absence of a symmetry 
protection.
In the model under consideration such a protection comes from the accidental $P_{LR}$-symmetry of the fermionic sector, which also implies that the 
vertex $hZ\gamma$ in this case is generated only by the $\rho$ exchange; the value of $c_{Z\gamma}$ is thus given by Eq.~(\ref{eq:cZgarho}).

\paragraph{Model 2} In the second model the composite fermions fill  one fundamental plus one antisymmetric representation
of $SO(5)$,   with $U(1)_X$ charge $+2/3$ and $-1/3$ respectively: 
\begin{equation}
\begin{split}
\ferm_5 & = \rep[1]{1}_{2/3} + \rep[2]{2}_{2/3} \\[0.1cm]
\ferm_{10} & = \rep[2]{2}_{-1/3} + \rep[1]{3}_{-1/3} + \rep[3]{1}_{-1/3} \, .
\end{split}
\end{equation}
As before, the Lagrangian can be divided into an elementary and a composite part plus a mixing term
\begin{equation}
\begin{split}
{\cal L}_{mix} = 
& \, \lambda^{10}_q \, \bar q_L P_q U(\pi) \ferm_{10} + \lambda^{5}_q \, \bar q_L P_q U(\pi) \ferm_{5}
+ \lambda^{10}_t \, \bar t_R P_t U(\pi) \ferm_{10}  \\
& + \lambda^{5}_t \, \bar t_R P_t U(\pi) \ferm_5  
+ \lambda_b \, \bar b_R P_b U(\pi) \ferm_{10} + h.c. \, .
\end{split}
\end{equation}
In this case the fermionic sector is  not in general $P_{LR}$ invariant, so both loops of composite fermions and the tree-level exchange of the $\rho$ 
can contribute to generate the $hZ\gamma$ vertex.
Using Eqs.~(\ref{eq:cZgarho}), (\ref{eq:final1loop}) and (\ref{eq:cZga})   we find
\begin{equation}
\label{eq:cZgamodel2}
\begin{split}
c_{Z\gamma} = & \, \frac{g^2}{2} \sin^2\!\theta \, (\alpha_{1L} - \alpha_{1R}) \\
& + \frac{g^2}{4} N_\chi \sin^2\!\theta  \, \Big[ |\zeta_{13}|^2 \left( C(m_{\rep[2]{2}}, m_{\rep[1]{3}}) - C(m_{\rep[1]{3}}, m_{\rep[2]{2}}) \right) \\
& \hspace{2.7cm} - |\zeta_{31}|^2 \left( C(m_{\rep[2]{2}}, m_{\rep[3]{1}}) - C(m_{\rep[3]{1}}, m_{\rep[2]{2}}) \right) \Big]\, .
\end{split}
\end{equation}
The shift to $Z\bar bb$ is suppressed for $\lambda_q^{10}$  small, since no effect can arise 
from the $P_{LR}$-preserving coupling~$\lambda_q^5$~\cite{Agashe:2006at}. As before, a small $\lambda_q^{10}/\lambda_q^{5}$ can arise naturally
from the RG flow of the full theory, and can explain the hierarchy between the top and bottom masses if $\lambda_t^{10} \ll \lambda_t^5 \simeq \lambda_b$.

\section{$S$ parameter from loops of fermionic resonances}
\label{sec:Sparameter}
\setcounter{equation}{0}

\begin{figure}[t]
\begin{center}
\includegraphics[width=0.28\linewidth]{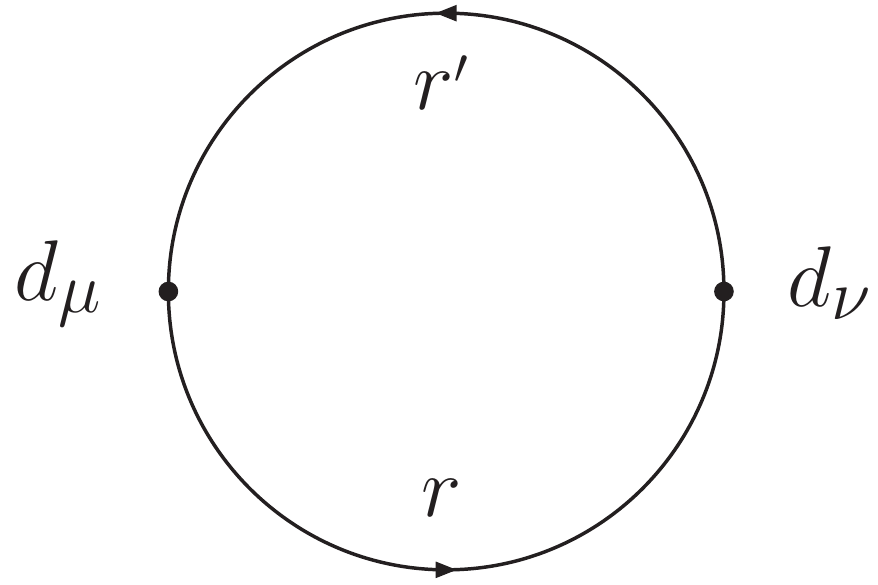}
\hspace{1.5cm}
\includegraphics[width=0.28\linewidth]{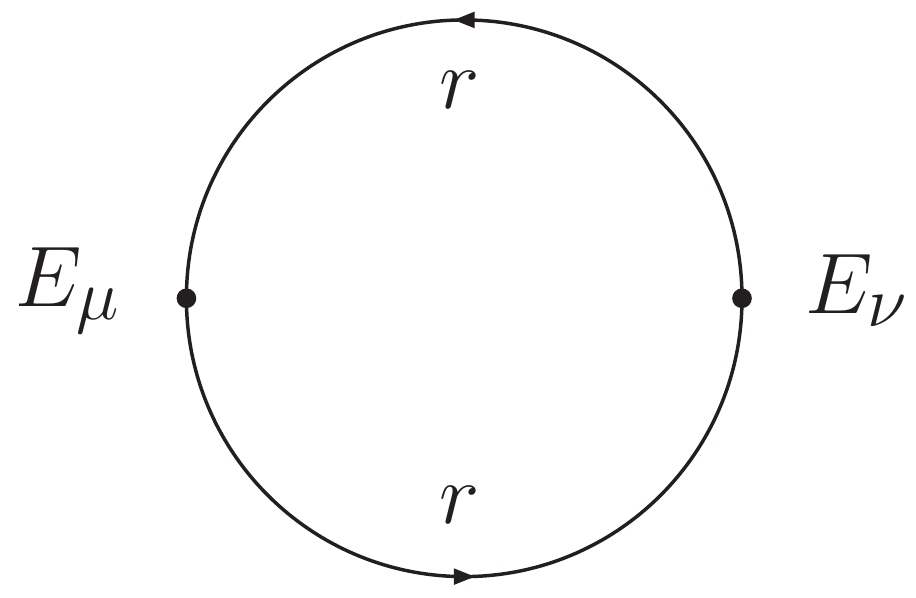}
\caption{\small 
One loop diagrams contributing to the $\langle d_\mu d_\nu \rangle$ (left) and $\langle E_\mu E_\nu \rangle$ (right) self-energies.
}
\label{fig:selfenergies}
\end{center}
\end{figure}

In the previous Section we have seen that loops of composite fermions generate the operator~$O_{d1}$ through the triangle diagram of Fig.~\ref{fig:triangle};
the value of the corresponding coefficient $c_{d1}$ is given by Eq.~(\ref{eq:cd12}).
Since $O_{d1}$ can be rewritten in terms of $O_3^+$ as in Eq.~(\ref{eq:opidentity}), it  contributes to the $S$ parameter. As implied by the Ward 
identity~(\ref{eq:WI}), the same contribution to $c_{d1}$, hence to $S$, can be derived by considering the $\langle d_\mu d_\nu \rangle$ self-energy diagram 
shown on the left of  Fig.~\ref{fig:selfenergies}, where two different $SO(4)$ representations of fermions circulate in the loop.
%
%
%
%
There is however an additional direct contribution to $O_3^\pm$ which comes from the $\langle E_\mu E_\nu \rangle$ self-energy diagram shown on the right 
of Fig.~\ref{fig:selfenergies}, where a single fermion representation $r$ appears in the loop.  Summing over $r$, we find
\begin{equation}
c_{3}^\pm = \frac{N_\chi}{8}  \sum_{r} (C_L[r]\pm C_R[r] ) A(m_r,m_r) \, ,
\end{equation}
where for the fundamental representation $C_L[\rep[2]{2}]= C_R[\rep[2]{2}] = 1$,  for the adjoint $C_R[\rep[1]{3}]=C_L[\rep[3]{1}] =2$
and $C_L[\rep[1]{3}]=C_R[\rep[3]{1}] =0$, while  $C_{L,R}[\rep[1]{1}]=0$.
The total contribution to the $S$ parameter from loops of composite fermions  is thus
\begin{equation}
\label{eq:Sgeneral}
\Delta S = -4\pi  N_\chi \sin^2\!\theta \left[ \sum_{r} ( C_L[r] + C_R[r] ) A(m_r,m_r) 
- \sum_{r,r^\prime}  |\zeta_{[r,r^\prime]}|^2  C[r,r'] A(m_r,m_{r^\prime})  \right]\, ,
\end{equation}
where we have conveniently defined $C[r,r'] = (1/2) ( l_{[r,r^\prime]}^L + l_{[r,r^\prime]}^R + l_{[r^\prime,r]}^L + l_{[r^\prime,r]}^R )$ and used the fact that the function
$A(m_r,m_{r'})$ is symmetric in its arguments.
For example, in the first model discussed in Section~\ref{sec:models} with fermions in the $\rep[1]{1}$ and $\rep[2]{2}$ of $SO(4)$ one has
\begin{equation}
\label{eq:Smodel1}
\begin{split}
\Delta S 
& =  - 8\pi N_\chi \sin^2\!\theta \left( A(m_{\rep[2]{2}},m_{\rep[2]{2}}) - |\zeta_{11}|^2 A(m_{\rep[2]{2}},m_{\rep[1]{1}})  \right) \\[0.2cm]
& =  \frac{2}{3}\, \frac{N_\chi}{\pi} \, \frac{v^2}{f^2}  \left( 1- |\zeta_{11}|^2 \right)  \log\!\bigg(\frac{\Lambda^2}{\bar m^2}\bigg)+ \text{finite terms}\, ,
\end{split}
\end{equation}
where in the second expression  $\bar m$ denotes an average mass 
and the finite terms include the proper ratios of fermion masses.
In the second model with fermions in the $\rep[2]{2}$, $\rep[1]{3}$ and $\rep[3]{1}$ we find
\begin{equation}
\label{eq:Smodel2}
\begin{split}
\Delta S 
& = -8\pi N_\chi \sin^2\!\theta \, \bigg[ A(m_{\rep[2]{2}},m_{\rep[2]{2}}) + A(m_{\rep[3]{1}},m_{\rep[3]{1}}) + A(m_{\rep[1]{3}},m_{\rep[1]{3}}) \\
& \phantom{= -8\pi N_\chi \sin^2\!\theta \, \bigg[ \, } -\frac{3}{2} |\zeta_{13}|^2 A(m_{\rep[2]{2}},m_{\rep[1]{3}}) 
-  \frac{3}{2} |\zeta_{31}|^2 A(m_{\rep[2]{2}},m_{\rep[3]{1}}) \bigg] \\[0.2cm]
& = \frac{1}{2}\, \frac{N_\chi}{\pi}\, \frac{v^2}{f^2}  \left( 2- |\zeta_{13}|^2  - |\zeta_{31}|^2 \right)  \log\!\bigg(\frac{\Lambda^2}{\bar m^2}\bigg)+ \text{finite terms}\, .
\end{split}
\end{equation}

From Eqs.(\ref{eq:Sgeneral})-(\ref{eq:Smodel2}) one can see that the $S$ parameter is in general logarithmically divergent, as expected on dimensional grounds.
The coefficient of the log  can be either positive or negative
depending on the value of the parameters $\zeta$.
Analogous  results were first obtained in the context of Technicolor theories in Ref.~\cite{Golden:1990ig} and later re-derived for 
$SO(4)/SO(3)$ Higgsless models by Ref.~\cite{Barbieri:2008zt}. 
More recently, the case of $SO(5)/SO(4)$ composite Higgs theories has been discussed in Ref.~\cite{Grojean:2013qca}.~\footnote{The same results hold
in 5-dimensional Holographic Higgs models. This can be most easily shown by solving the bulk dynamics and deriving the holographic action on the
boundary where the elementary fields live, see for example Ref.~\cite{Panico:2007qd}. 
In absence of boundary terms, the 4D holographic action for the fermions has the  CCWZ form with $\zeta =1$.
This is indeed the reason why previous 1-loop calculations  in the context of 5D models found a finite $S$ parameter, see for example Refs.~\cite{5DSparameter}.
Values $\zeta \not =1$ can be obtained by introducing the boundary term $F^{\hat a}_{\mu 5} \bar\psi T^{\hat a} \gamma^\mu \psi$, since
by using the  equations of motion in the bulk it follows $F^{\hat a}_{\mu 5} \propto d_\mu^{\hat a}(\pi)$.
}
%
A simple way to understand why the log divergence vanishes if the parameters $\zeta$ are equal to~1 is by noticing
that in this limit the Lagrangian~(\ref{eq:Lfermion}) can be rewritten, through a field redefinition, as the Lagrangian
of a two-site model where the Higgs couplings to the composite fermions are non-derivative.~\footnote{The same observation was  recently made by 
Ref.~\cite{Grojean:2013qca}.}
This implies, by simple inspection of the relevant one-loop  diagrams, that the $S$ parameter
is finite in this case. For completeness we report in Appendix~\ref{sec:fccwz} a short discussion on the connection between the CCWZ Lagrangian~(\ref{eq:Lfermion}) and
that of the two-site model.

The fact that the overall sign of $S$ is controlled by the coefficients $\zeta$ and can be negative is more clearly understood 
by considering  the dispersion relation obeyed by $S$~\cite{Orgogozo:2012ct}:
\begin{equation}
\label{eq:disprel}
S= 4\pi \sin^2\!\theta \int \!\frac{ds}{s} \left[ \rho_{LL}(s) + \rho_{RR}(s) - 2 \rho_{BB}(s) \right]\, ,
\end{equation}
where $\rho_{LL}$, $\rho_{RR}$ and $\rho_{BB}$ are  the spectral functions respectively of two unbroken ($SU(2)_L$ and $SU(2)_R$) and broken ($SO(5)/SO(4)$) 
conserved currents of the strong sector.  The  definition of the spectral function $\rho(s)$ and the expression of the currents is reported in 
Appendix~\ref{sec:spectralfunction} for completeness.
From Eq.~(\ref{eq:disprel}) and from the positivity of each individual spectral function, it is clear that a negative $S$
can occur  if $\rho_{BB}$  is sufficiently large.
The leading contribution of the  fermions to the spectral functions can be easily computed from the diagrams shown in Fig.~\ref{fig:spectralfunct}. We find:
%
\begin{figure}[t]
\begin{center}
\includegraphics[height=0.16\linewidth]{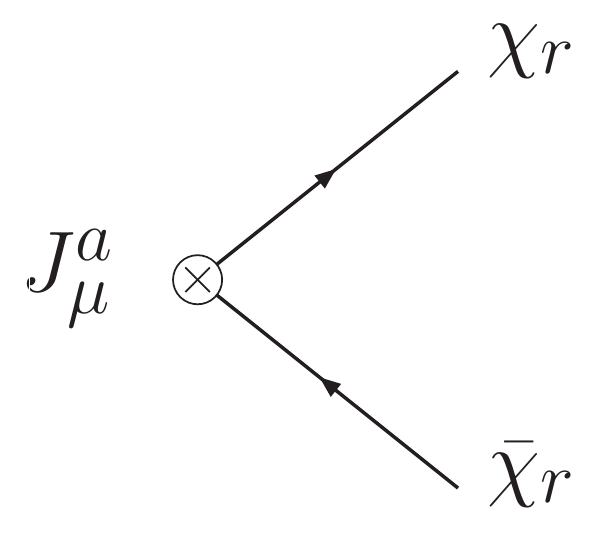}
\\[0.5cm]
\includegraphics[height=0.16\linewidth]{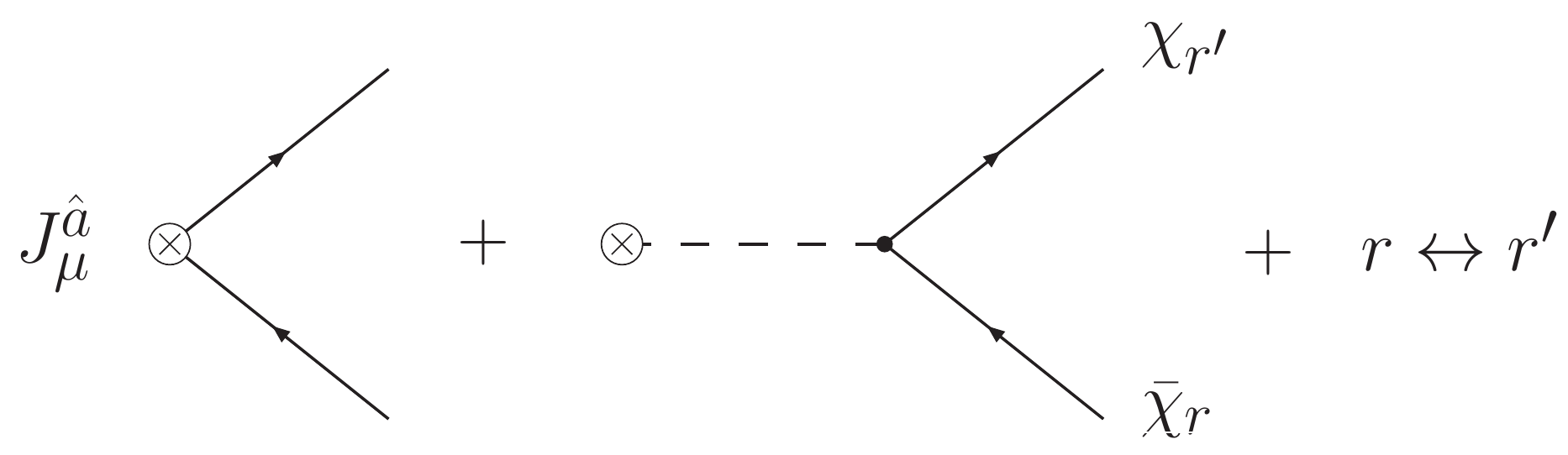}
\caption{\small 
Two-particle contribution to the spectral function of unbroken (upper row) and broken (lower row) currents from the composite fermions. The dashed line 
denotes the propagator of a NG boson.
}
\label{fig:spectralfunct}
\end{center}
\end{figure}
%
%
\begin{equation}
\begin{split}
\rho_{LL,RR}(q^2) & = \frac{1}{12\pi^2} \sum_r C_{L,R}[r] \,\lambda(q^2 , m_r , m_{r}) \\[0.2cm]
\rho_{BB}(q^2) & = \frac{1}{24\pi^2}  \sum_{r,r'} |\zeta_{[r,r^\prime]}|^2 C[r,r^\prime] \,\lambda(q^2 , m_r , m_{r'}) \, ,
\end{split}
\end{equation}
where we have defined
\begin{equation}
\begin{split}
\lambda(q^2 , m_r , m_{r'}) = & \, \left( 1- \frac{(m_r - m_{r^\prime})^2}{q^2} \right) \left( 1+ \frac{(m_r + m_{r^\prime})^2}{2 q^2} \right) \\[0.1cm]
& \times \sqrt{\left(1+ \frac{m_r^2-m_{r^\prime}^2}{q^2} \right)^2 - 4\, \frac{m_r^2}{q^2} }\, .
\end{split}
\end{equation}
By inserting these expressions into the dispersion relation (\ref{eq:disprel}), one re-obtains the result of Eq.~(\ref{eq:Sgeneral}).
Since $\rho_{BB}$ is proportional to $|\zeta_{[r,r^\prime]}|^2$, it is clear that for sufficiently large $|\zeta_{[r,r^\prime]}|$ the $S$ parameter
will become negative.

\section{Numerical Results and Discussion}
\label{sec:numericalanalysis}
\setcounter{equation}{0}

In this paper we have focused on the virtual effects due to purely composite states. The Higgs decay rate to $Z\gamma$ and the $S$ parameter
are two low-energy observables extremely sensitive to such effects.~\footnote{We are particularly grateful to John Terning  for drawing our attention to the possibility of correlation between these two effects.} It is well know that the tree-level contribution to $S$ from spin-1 resonances
is large and poses tight constraints on the scale of compositeness. We have seen that the exchange of $\rho^{L}$ and $\rho^R$ generates the effective
interaction $hZ\gamma$ also  at  tree level, provided their masses and couplings are not $P_{LR}$ symmetric.
This leads to a correction to the $h\to Z\gamma$ decay rate that is potentially larger than that due to the $O(v^2/f^2)$ shifts in the tree-level Higgs couplings from
the  non-linear $\sigma$-model Lagrangian.  This is the case  unless the coefficients of the operators $Q_{1L}$ and $Q_{1R}$ are loop suppressed, as  happens
for example in Holographic  Higgs theories.
The contribution from fermionic resonances arises  at the 1-loop level, and can be numerically large. The main reason for this is that loops of
pure composites are sensitive to the multiplicity of states arising from the strong dynamics. In particular all the composite fermion species,
including the partners of SM light quarks and leptons,
will circulate in the loop regardless of how strongly mixed with the elementary  fermions they are. 
The multiplicity factor $N_\chi$  can then partly compensate for the one-loop suppression, giving large shifts to both the $S$ parameter 
and the $h\to Z\gamma$ rate.~\footnote{One might worry that a large multiplicity factor $N_\chi$ could invalidate the perturbative expansion.
However, 
the light Higgs mass already indicates that composite fermions must be somewhat more weakly coupled than other resonances, see for example 
Refs.~\cite{Panico:2012uw,DeSimone:2012fs}. 
With $\sim 1\,$TeV fermion masses and $f=500-800\,$GeV, for example, the coupling strength $g_* = M/f$ is sufficiently small to allow a 
perturbative expansion controlled by the loop  parameter $N_\chi (g_*^2/16\pi^2)$. 
}
%
\begin{figure}[t]
\begin{center}
\includegraphics[height=5.9cm]{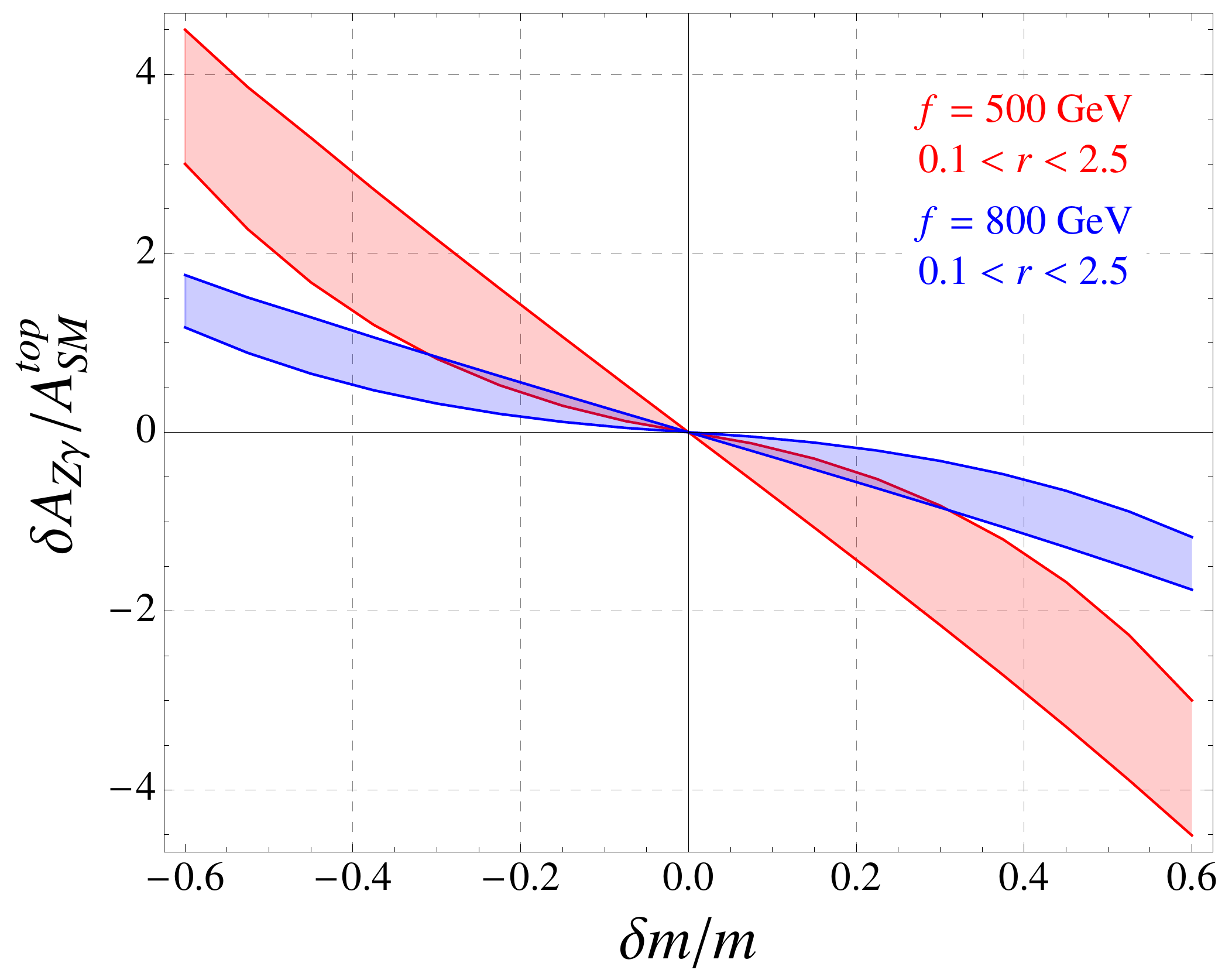}
\hspace{0.4cm}
\includegraphics[height=5.9cm]{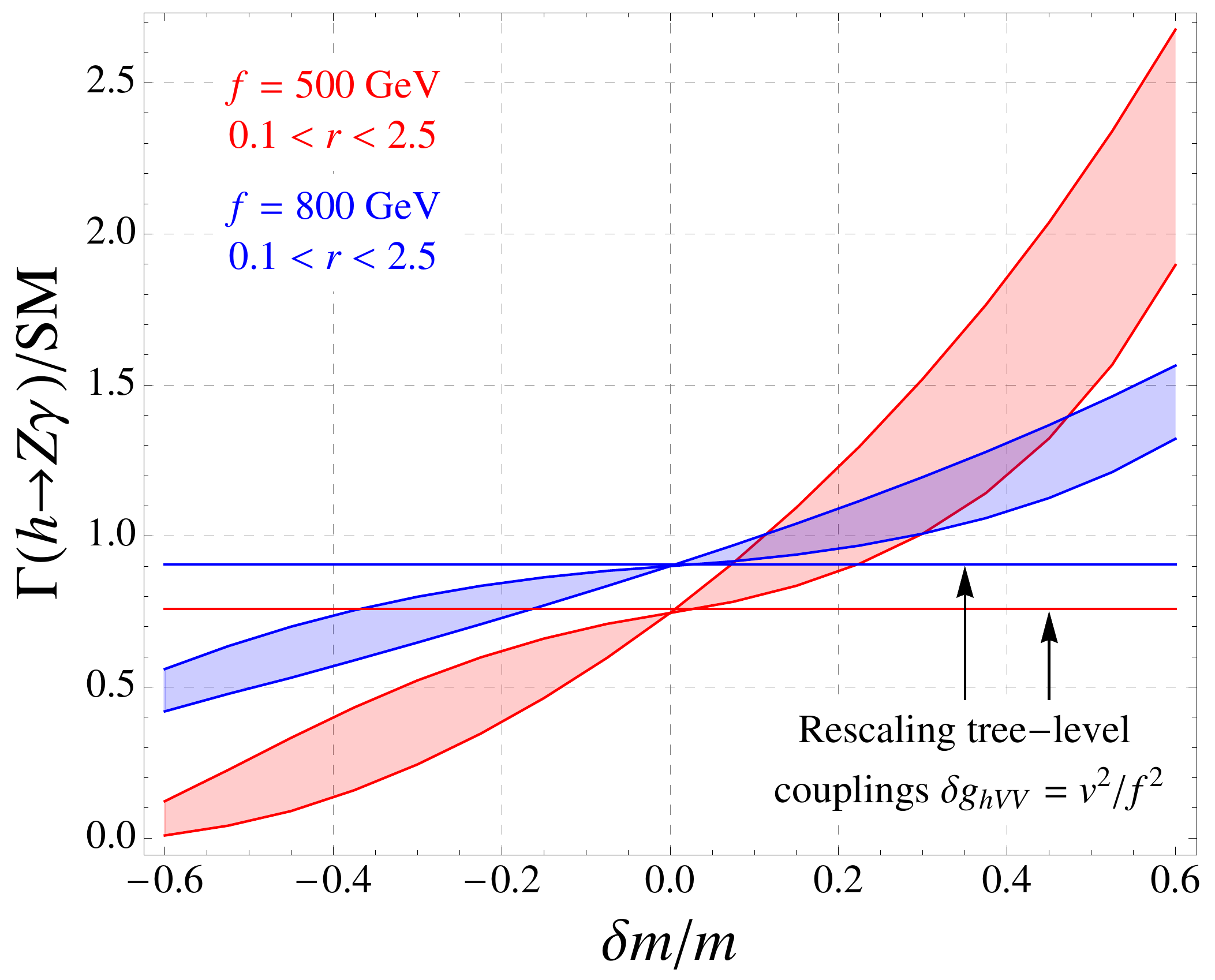}
\caption{\small 
Left plot: shift of the $h\to Z\gamma$ decay amplitude in units of the SM top contribution,
$\delta A/A^{top}_{SM}$, in the second model of Section~\ref{sec:models} as a function of the $LR$ mass splitting. 
Right plot: total decay rate of $h\to Z\gamma$  normalized to its SM value, $\Gamma/\Gamma_{SM}$,  in the same model. 
The left plot assumes one family of colored  fermions ($N_\chi =3$), while the right plot assumes three degenerate families of composites ($N_\chi =9$).
The horizontal lines indicate the value obtained by  including only the effect of the modified tree-level Higgs couplings.
}
\label{fig:numericalZgamma}
\end{center}
\end{figure}
%

To illustrate the size of the effects we have been discussing, the left plot of Fig.~\ref{fig:numericalZgamma} shows the shift to the $h\to Z\gamma$ decay amplitude 
in units of the SM top contribution, $\delta A/A^{top}_{SM}$, due to one family of colored fermions (composite quarks)
transforming as a $\mathbf{10}+\mathbf{5}$ of $SO(5)$ (second model of Section~\ref{sec:models} with $N_\chi =3$).
As discussed in Section~\ref{sec:fermloops}, the correction comes entirely from the~$\mathbf{10}$, hence the relevant parameters 
 are the following:  the scale of compositeness~$f$,  the coefficients $\zeta_{13}$, $\zeta_{31}$, and  two ratios of masses which we conveniently define to be
$\delta m/m \equiv (m_{\rep[3]{1}} -m_{\rep[1]{3}})/(m_{\rep[3]{1}}+m_{\rep[1]{3}})$ and $r \equiv m_{\rep[2]{2}}/(m_{\rep[3]{1}}+m_{\rep[1]{3}})$.
For simplicity we fix $\zeta_{13}=\zeta_{31} =1$, so that the amount of $P_{LR}$ breaking is fully controlled by $\delta m/m$. The plot  
shows the relative shift $\delta A/A^{top}_{SM}$  as a function of $\delta m/m$ for two representative values $f = 500\,$GeV and  $f = 800\,$GeV.
The red and blue bands are obtained by  varying $r$ in the interval $0.1 < r < 2.5$.
By rescaling $\zeta_{13}$ and $\zeta_{31}$ by a common factor $\zeta$, $\delta A$ goes like~$\zeta^2$, though even without such an enhancement  we see that
shifts of several times the SM top amplitude are possible for large mass splittings.  
The right plot of Fig.~\ref{fig:numericalZgamma} shows the total decay rate normalized to its SM value, this time for three degenerate families of colored fermions
(second model of Section~\ref{sec:models} with $N_\chi =9$).
The horizontal lines indicate the
value obtained by  including only the effect of the modified tree-level Higgs couplings discussed above.
Since in the SM the $W$ loop contribution largely dominates that of the top quark,
the  effect from the modified tree-level couplings is a suppression of the decay rate by a factor $(g_{WWh}/g_{WWh}^{SM})^2 = (1-v^2/f^2)$.
The correction from the 1-loop exchange of composite fermions is included in addition to this effect, and can further suppress or  enhance the decay rate
depending on the sign of the mass splitting $\delta m/m$.

\begin{figure}[t]
\begin{center}
\includegraphics[width=0.5\linewidth]{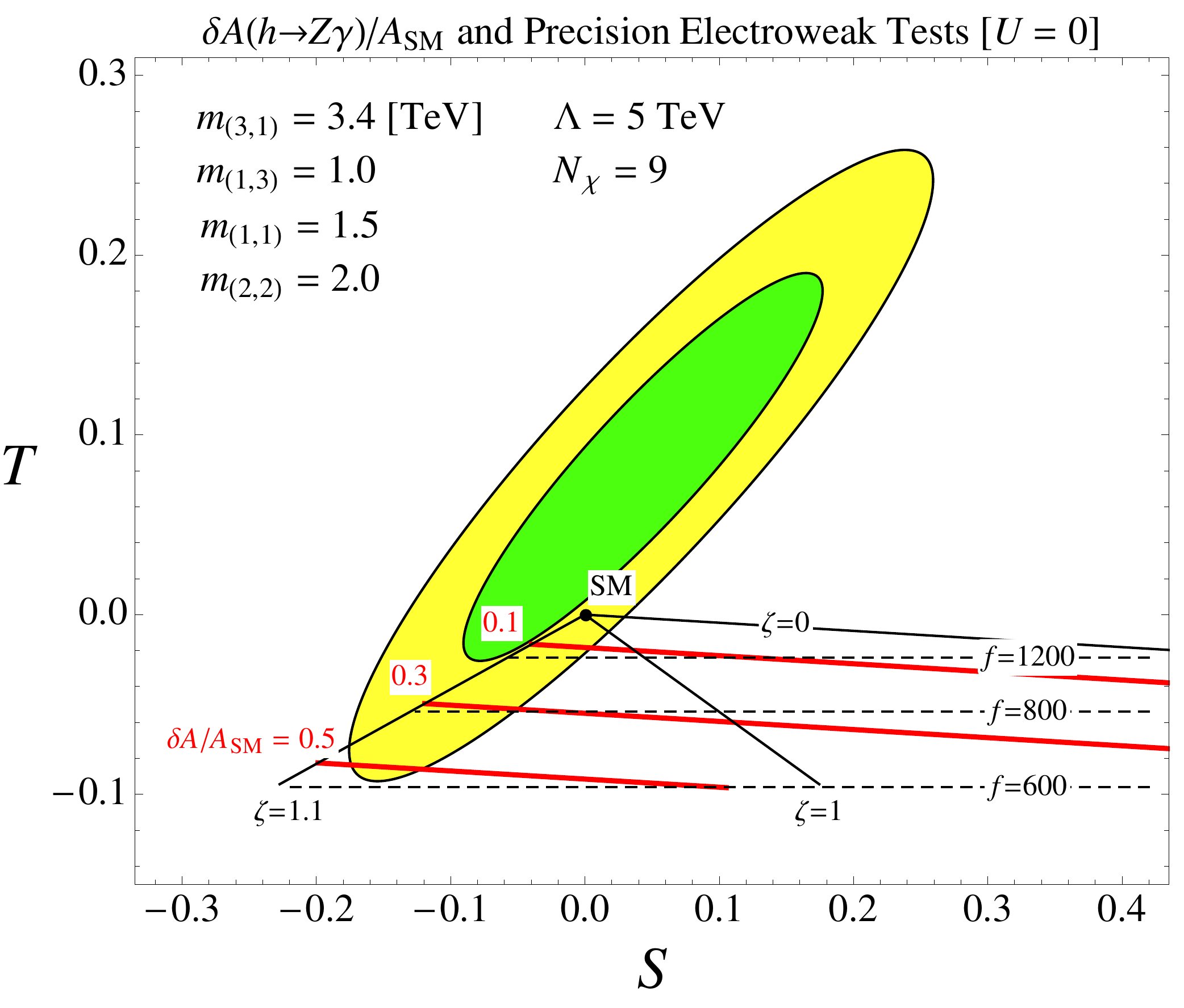}
\caption{\small 
Region spanned in the plane $(S,T)$ when varying $f$ and $\zeta=\zeta_{13} = \zeta_{31} =\zeta_{11}$ in the second model of Section~\ref{sec:models}, as due
to $\Delta S$ from loops of  composite fermions and to the IR correction to $S$ and $T$ from modified Higgs couplings.
The green  and yellow areas indicate the regions with 68\% and 95\% probability~\cite{Baak:2012kk}. Dashed (solid black) lines indicate the trajectories
of fixed $f$ ($\zeta$). The thicker solid red lines indicate the isocurves of constant $h\to Z\gamma$ decay amplitude.
}
\label{fig:STplane}
\end{center}
\end{figure}

It is interesting to derive the contribution to the $S$ parameter in this model and analyze the impact of a sizable correction to the
$h\to Z\gamma$ decay rate on the EWPT. This is illustrated by Fig.~\ref{fig:STplane}  in the $(S,T)$ plane.~\footnote{The probability
contours  have been derived by using the fit on $(S,T)$ performed by the GFitter collaboration~\cite{Baak:2012kk}. Similar results are obtained by using the
more recent analysis of Ref.~\cite{Ciuchini:2013pca}.}
The plot shows the region spanned 
by varying $f$ and $\zeta \equiv \zeta_{13} = \zeta_{31} =\zeta_{11}$  due to the  IR correction to $S$ and $T$ from modified Higgs couplings
and to the 1-loop correction to $S$ from three degenerate families of composite fermions (Eqs.~(\ref{eq:Smodel1}) and (\ref{eq:Smodel2}) with $N_\chi =9$).
%
%
We have fixed the cutoff scale to $\Lambda = 5\,$TeV, and have chosen the following spectrum of
composite masses: $m_{\rep[1]{1}} = 1.5\,$TeV,  $m_{\rep[2]{2}} = 2.0\,$TeV,  $m_{\rep[3]{1}} = 3.4\,$TeV,  $m_{\rep[1]{3}} = 1.0\,$TeV, so that $r =0.45$ and $\delta m/m=0.55$.
Even in the absence of additional contributions to $T$, the correction  to $S$ from loops of composite fermions can  compensate the shift due to the  modified couplings
of the Higgs to the SM vector bosons and bring the theory point back into the $95\%$ probability contour.
For example, for $f=800\,$GeV (i.e. $(v/f)^2 \simeq 0.09$) one has $\Delta S \simeq 0.83\, (1-\zeta^2)$ from composite fermions, so that $\zeta \sim 1.1$ gives
 $\Delta S \sim -0.2$ as required to offset the IR shift.
Correspondingly, the correction to the $h\to Z\gamma$ rate is sizable and of order $70\%$ of the SM value.
In general, the 1-loop contribution to $S$ is large and only values $\zeta \simeq 1$ are viable.
The fact that EWPT select a narrow range of $\zeta$ is directly relevant for the experimental searches of the fermionic resonances, since $\zeta$ controls
their single production~\cite{DeSimone:2012fs}.~\footnote{We thank Minho Son for drawing our attention to this point.}
The exact allowed range depends however on possible additional  contributions to $S$ and $T$.
For example, for $g_{\rho_L} =3$ and $f=800\,$GeV the contribution to $S$ from a $\rho^L$ is $\Delta S \simeq 0.13$ (see Eq.~(\ref{eq:Sparameter})), which increases 
the preferred value of $\zeta$ by only a $5-10\%$, which in turn corresponds to an increase of the $h\to Z\gamma$ rate by $10-20\%$.

The tuning required to comply with the EW precision tests can be alleviated if an additional positive contribution to $T$ is present.
This can arise from loops of fermionic resonances, as recently discussed by Ref.~\cite{Grojean:2013qca}; see also Refs.~\cite{Pomarol:2008bh}.
Unlike  the $S$ parameter, however, $T$ is generated only if the custodial invariance of the strong dynamics is broken, and therefore
no correction can come from purely composite states. In theories with partial compositeness and flavor anarchy of the strong sector, the
leading contribution arises from loops of elementary top quarks. For example, if $t_R$ mixes with a composite singlet of $SU(2)_L \times SU(2)_R$, as in model~1
of Section~\ref{sec:models},
the only breaking of  custodial symmetry in the fermionic sector comes from $\lambda_q$.
As a spurion analysis shows~\cite{Giudice:2007fh}, one needs four powers of $\lambda_q$ to generate $T$, 
which implies a finite result (i.e. independent of the cutoff scale $\Lambda$). A subleading contribution comes from loops of spin-1 resonances and 
elementary hypercharge vector bosons. In this case the breaking of custodial symmetry comes from the hypercharge coupling, and  two powers of $g^\prime$
are sufficient to generate $T$. 
Table~\ref{tab:NDAestimates} summarizes the naive estimates of the corrections to $\hat S \equiv (\alpha_{em}/4\sin^2\!\theta_W) S$
and $\hat T \equiv \alpha_{em} T$.
%
\begin{table}[t]
\begin{center}
{\small
\begin{tabular}{ll|cc}
 && $\Delta \hat S$ & $\Delta \hat T$  \\
\hline 
&&& \\[-0.3cm]
 \multirow{2}{1cm}[-0.55cm]{UV} & $\chi$  & $\displaystyle  N_\chi \frac{g^2}{16\pi^2} \left(\frac{v^2}{f^2}\right) \log\!\left(\frac{\Lambda}{M}\right)$ 
                                                & $\displaystyle N_c \frac{\lambda_q^2}{16\pi^2} \left(\frac{v^2}{f^2}\right) \frac{\lambda_q^2}{g_*^2}  $\\[0.8cm]
                                & $\rho$ & $\displaystyle \left(\frac{v^2}{f^2}\right)  \left(\frac{g^2}{g_*^2}\right)$ 
                                                & $\displaystyle \frac{g'^{2}}{16\pi^2} \left(\frac{v^2}{f^2}\right) \log\!\left(\frac{\Lambda}{M}\right)$ \\[1cm]
 \multirow{2}{1cm}[-0.55cm]{IR}   & NGB     & $\displaystyle \frac{g^2}{16\pi^2} \left(\frac{v^2}{f^2}\right) \log\!\left(\frac{M}{\mu}\right)$ 
                                                & $\displaystyle \frac{g'^2}{16\pi^2} \left(\frac{v^2}{f^2}\right) \log\!\left(\frac{M}{\mu}\right)$ \\[0.8cm]
                                & top       & $\displaystyle N_c \frac{g^2}{16\pi^2} \left(\frac{v^2}{f^2}\right) \frac{\lambda_q^2}{g_*^2} \log\!\left(\frac{M}{\mu}\right)$
                                                & $\displaystyle N_c \frac{y_t^2}{16\pi^2} \left(\frac{v^2}{f^2}\right) \frac{\lambda_q^2}{g_*^2} \log\!\left(\frac{M}{\mu}\right)$
\end{tabular}
}
\\[0.3cm]
\caption{\small Naive estimates of the UV corrections (from composite fermions $\chi$ and spin-1 resonances~$\rho$) and IR
corrections (from NG bosons and the top quark) to $\hat S \equiv (\alpha_{em}/4\sin^2\!\theta_W) S$ and $\hat T \equiv \alpha_{em} T$.
}
\label{tab:NDAestimates}
\end{center}
\end{table}
%
As before,  $g_* \sim M/f\sim g_\rho$ denotes the coupling strength of the composite states and  $M$ their mass scale.
The first two lines show the corrections discussed above that arise from the exchange of composite fermions and spin-1 resonances. These 
are short-distance effects at the scale $M$, which in the language of the Higgs effective Lagrangian correspond 
to threshold corrections to the Wilson coefficients $\bar c_W + \bar c_B$ and $\bar c_T$; see Ref.~\cite{Contino:2013kra}.
There are however additional contributions which are generated by the exchange of light SM fields below $M$ and are thus associated to the RG evolution of the Wilson 
coefficients down to IR scales $\mu\approx m_Z$.
The largest corrections arise from loops of NG bosons (i.e.  longitudinally polarized $W$ and $Z$ and the Higgs boson) and of  top quarks,
and correspond to the RG evolution of $\bar c_W + \bar c_B$ and $\bar c_T$ due to  $\bar c_H$ and $\bar c_{H\psi}$, respectively \cite{Contino:2013kra}.
Their naive estimates are reported in the last two lines of Table~\ref{tab:NDAestimates}.
Loops of transverse gauge bosons also lead to IR corrections which are  subleading. For example, as recently pointed out by the authors of 
Ref.~\cite{Falkowski:2013dza}, 1-loop diagrams featuring one insertion of the effective $hZ\gamma$ vertex (induced by the operator $O_{HW}-O_{HB}$) give 
a correction to the $\hat S$ parameter of order 
\begin{equation}
\Delta \hat S \sim \left( \frac{g^2}{16\pi^2} \right)^2 \left(\frac{v^2}{f^2}\right) \log\!\left( \frac{M}{\mu} \right)^2\, .
\end{equation}
Although directly linked to $h\to Z\gamma$, this is a  two-loop EW effect which is parametrically subleading compared to other IR effects  and numerically
smaller than the UV corrections from pure composite fermions (see Table~\ref{tab:NDAestimates}).

\vspace{0.5cm}
We briefly summarize our findings with the following conclusions:
\begin{itemize}
\item The decay mode $h \to Z \gamma$, unlike other loop-mediated processes of a Nambu-Goldstone composite Higgs, is subject to NP corrections that are not suppressed by the Goldstone symmetry itself.  While new contributions to the $hgg$ and $h \gamma \gamma$ contact interactions of the effective Lagrangian are typically (and observably) small, a highly nonstandard $hZ \gamma$ interaction is possible and consistent with the symmetry that is assumed to be responsible for stabilizing the weak scale.  
\item Generating a large $hZ \gamma$ interaction in the absence of significant breaking of the Goldstone symmetry relies on the intervention of states arising from a strong sector that breaks a left-right symmetry, $P_{LR}$.  Provided this breaking is mediated in a suppressed way to the $Zb \bar b$ coupling, as in the case of the models presented above, enhancements of $h \to Z \gamma$ remain phenomenologically viable.
\item There are two operators contributing to the $S$ parameter that are  closely related to those governing  $h \to Z \gamma$, and a naive prediction would be for a tight correlation between these two observables.  However, the composite Higgs can couple to fermions through interactions that contribute only to the two-point function of two broken currents, allowing an offsetting (negative) contribution to $S$ such that again the viability of large corrections in $h\to Z \gamma$ is retained.
\end{itemize} 
In this paper we have highlighted the anatomy of the $h \to Z \gamma$ channel, one in which a composite Higgs might naturally interact in a novel way that could help shed light on its origins in the absence of other, more obvious, clues.  As such, this channel deserves our full attention in the continuation of Higgs study at the LHC.

\section*{Acknowledgments}

We would like to thank 
David Marzocca,
Riccardo Rattazzi,
Slava Rychkov,
Marco Serone,
Minho Son,
John Terning,
and
Enrico Trincherini
for useful discussions.
We  also thank Leandro~Da~Rold and Eduardo~Pont\'on for pointing out a few typos in the first version of the paper. 
The work of A.A., R.C. and J.G. was partly supported by the ERC Advanced Grant No.~267985 
\textit{Electroweak Symmetry Breaking, Flavour and Dark Matter: One Solution for Three Mysteries (DaMeSyFla)}.

\appendix

\section{Formulas for the $h\to Z\gamma$ decay rate}
\label{app:decayrate}
\setcounter{equation}{0}

We collect here the formulas useful for the calculation of the decay rate $h\to Z\gamma$.
The partial width is given by
\begin{equation}
\label{eq:decayrate}
\Gamma(h\to Z\gamma)=\frac{1}{32\pi} \frac{m_h^3}{v^2}\l(1-\frac{m_Z^2}{m_h^2}\r)^3 \l|A\r|^2\, ,
\end{equation}
where $A$ is the total decay amplitude. The SM contribution arises from loops of $W$ vector bosons and fermions:
\begin{equation}
\label{eq:SMamplitudes}
\begin{split}
A_{SM}= & A_F+A_W \\[0.5cm]
A_F = & -\frac{\alpha_{em}}{\pi} \sum_f N_{cf} Q_f  \frac{\l(T_f^{3L}-2 Q_f \sin^2 \theta_W \r)}{\sin\theta_W \cos\theta_W} 
\l[ I_1(\tau_f,\lambda_f)-I_2(\tau_f,\lambda_f)\r] \\[0.25cm]
A_W= & -\frac{\alpha_{em}}{2\pi} \cot\theta_W  \, \bigg\{  4(3-\tan^2\theta_W)I_2(\tau_W,\lambda_W) \\
          & \hspace{2.8cm} +\l[\l(1+\frac{2}{\tau_W}\r)\tan^2\!\theta_W -\l(5+\frac{2}{\tau_W}\r)\r] I_1(\tau_W,\lambda_W) \bigg\}\, ,
\end{split}
\end{equation}
where $N_{cf}$ and $Q_f$ are respectively the number of color and the electromagnetic charge of the fermion $f$, and we have defined
\begin{equation}
\tau_f\equiv \frac{4 m_f^2}{m_h^2}\, , 
\qquad \lambda_f=\frac{4 m_f^2}{m_Z^2}\, ,
\qquad \tau_W=\frac{4 m_W^2}{m_h^2}\, ,
\qquad \lambda_W=\frac{4 m_W^2}{m_Z^2}\, .
\end{equation}
The loop functions are  equal to 
\begin{equation}
\label{eq:SMloopfuncts}
\begin{split}
I_1(a,b) & =\frac{a b}{2(a-b)}+\frac{a^2 b^2}{2(a-b)^2} \, (f(a)-f(b)) +\frac{a^2b}{(a-b)^2} \, (g(a)-g(b)) \\[0.15cm]
I_2(a,b) & =-\frac{ab}{2(a-b)}\l(f(a)-f(b)\r) \\[0.4cm]
g(\tau) & =
\begin{cases}
\sqrt{\tau-1} \,\arcsin (1/\sqrt{\tau}), &  \tau \geq 1 \\[0.1cm]
\frac{1}{2}\sqrt{1-\tau}\l[\log( \eta_+/\eta_-)-i \pi\r],  & \tau<1 
\end{cases} \\[0.4cm]
f(\tau) & =
\begin{cases}
\l[\arcsin (1/\sqrt\tau) \r]^2, & \tau \geq 1\\[0.1cm]
-{\displaystyle \frac{1}{4}} \l[\log (\eta_+/\eta_-)-i\pi\r]^2, & \tau<1\, ,
\end{cases}
\end{split}
\end{equation}
where $\eta_{\pm}\equiv (1\pm \sqrt{1-\tau})$.
In the limit in which the New Physics effect 
can be parametrized by the effective Lagrangian of Eq.~(\ref{eq:ZgammaOP}), the contribution to the decay amplitude is given by
\begin{equation}
A_{NP}=-2 c_{Z\gamma}\, .
\end{equation}
Numerically evaluating the SM contribution one finally obtains~\cite{Contino:2013kra}:
\begin{equation}
\frac{\Gamma(h\rightarrow Z\gamma)}{\Gamma(h\rightarrow Z\gamma)_{SM}}\simeq\left|1+0.01 \frac{4\pi}{\alpha_{em} \cos \theta_w} \, c_{Z\gamma}\right|^2
\simeq 1+0.02 \frac{4\pi}{\alpha_{em} \cos \theta_w} \, c_{Z\gamma}\, .
\end{equation}

\section{Relation between different bases of operators}
\label{app:opbases}
\setcounter{equation}{0}

In this Appendix we discuss the relations between our basis of operators (\ref{eq:CCWZbasis}) and those adopted in Ref.~\cite{Contino:2011np} (the CMPR basis
for short) and Ref.~\cite{Giudice:2007fh} (the SILH Lagrangian).

The CMPR list of CP-even operators is given by
\begin{equation}
\begin{split}
\cO_3 & = \Tr \big[ \big(E_{\mu\nu}^L \big)^2 - \big(E_{\mu\nu}^R \big)^2 \big]\\
\cO_4^\pm & = \Tr\!\left[\left( f_{\mu\nu}^L \pm f_{\mu\nu}^R \right) i [d^\mu,d^\nu] \right]\\
\cO_5^+ & =\Tr \!\left( (f_{\mu\nu}^-)^2 \right) \\
\cO_5^- &= \Tr\!\l[(f_{\mu\nu}^L)^2 - (f_{\mu\nu}^R)^2\r]\, ,
\end{split}
\end{equation}
plus other two operators, $\cO_{1} = O_1$ and $\cO_{2} = O_2$, whose expansion in terms of NG bosons starts at dimension 8.
Here $f_{\mu\nu}^{L,R,}$ and $f_{\mu\nu}^{-}$ are the dressed field strengths along the $SU(2)_L\times SU(2)_R$ and $SO(5)/SO(4)$ directions~\cite{Contino:2011np}.
We can relate the CMPR set to our basis by using the  identity
\begin{equation}
f_{\mu\nu}^L +f_{\mu\nu}^R = E^L_{\mu\nu} + E^R_{\mu\nu} + i[d_\mu, d_\nu] \, ,
\end{equation}
which holds for $SO(5)/SO(4)$. We find:
\begin{equation}
\begin{split}
\cO_3 &=O_3^- \\
\cO_4^+ &=O_4^+-\frac{1}{2}O_2+\frac{1}{2}O_1\\
\cO_4^- &=O_4^--O_5\\
\cO_5^+ & =O_3^++2O_4^++\frac{1}{2}O_1 -\frac{1}{2} O_2\\
\cO_5^- &=O_3^-+ 2O_4^- - O_5\, .
\end{split}
\end{equation}
The advantage of our basis over the CMPR one is that the connection to the SILH Lagrangian is more straightforward, since only four operators 
start at  dimension 6 when expanded in powers of the NG bosons. Also, only one operator gives a   $hZ\gamma$ contact interaction.

At the dimension-6 level, the connection between our operators and those of the SILH Lagrangian
is given  by
\begin{equation}
\begin{split}
-\frac{4f^2}{m_W^2} O_4^\pm & = O_{HW} \pm O_{HB}+ \dots\\[0.15cm]
- \frac{ f^2}{m_W^2}O_3^\pm & = O_W \pm O_B+\cdots 
\end{split}
\end{equation}
where the dots stand for dimension-8 terms.  The SILH operators are defined in Eqs.~(\ref{eq:SILHOP}) and (\ref{eq:SILHOP2}).

\section{Loop functions}
\label{app:loopfuncts}
\setcounter{equation}{0}

We collect here the expression of the loop functions $A(m_r,m_{r^\prime}),B(m_r,m_{r^\prime}),C(m_r,m_{r^\prime})$ defined in Eq.~(\ref{eq:Idec}):
\begin{align}
\begin{split}
A(m_r,m_{r'}) = \,
&\frac{1}{24 \pi ^2}\bigg[ 
-\log \left(\frac{\Lambda ^4}{m_r^2 m_{r'}^2}\right)
-\frac{m_r m_{r'} \left(-4 m_rm_{r'}+3 m_{r'}^2+3 m_r^2\right)}{\left(m_r^2-m_{r'}^2\right)^2}  \\[0.2cm]
&\hspace{1.15cm} +\frac{\left(6 m_r^3 m_{r'}^3-3 m_r^2 m_{r'}^2 \left(m_{r'}^2+m_r^2\right)+m_{r'}^6+m_r^6\right)  
                         \displaystyle \log\!\left(m_r^2/m_{r'}^2 \right)}{\left(m_r^2-m_{r'}^2\right)^3} \bigg] \, , 
\end{split} 
\\[0.3cm]
\begin{split}
B(m_r,m_{r'}) = \, 
& \frac{1}{24 \pi ^2} \, \frac{1}{\left(m_r^2-m_{r'}^2\right)^3} \, \times \\[0.1cm]
& \times \bigg[  2 m_{r'}^4 \left(m_{r'}^2-3 m_r^2\right) \log\!\left(\frac{m_{r'}^2}{\Lambda ^2}\right)-2 m_r^4
               \left(m_r^2-3 m_{r'}^2\right) \log\!\left(\frac{m_r^2}{\Lambda ^2}\right) \\[0.05cm]
& \hspace{0.7cm} -7 m_r^4 m_{r'}^2+7 m_r^2 m_{r'}^4-m_{r'}^6+m_r^6 \bigg] \, ,
\end{split}
\\[0.3cm]
\label{eq:functionC}
\begin{split}
C(m_r,m_{r'}) = \, 
& \frac{1}{24\pi^2} \, \frac{1}{(m_r^2-m_{r'}^2)^3}\, \times \\[0.1cm]
& \times \bigg[ 2 \left(3 m_r^2 m_{r'}+3 m_r m_{r'}^2-2 m_{r'}^3-3 m_r^3\right) m_{r'}^3 \log\!\left(\frac{m_{r'}^2}{\Lambda ^2}\right) \\[0.05cm]
& \hspace{0.6cm} +2 m_r \left(-6 m_r^3 m_{r'}^2+3 m_r^2 m_{r'}^3+3 m_r m_{r'}^4-3 m_{r'}^5+2 m_r^5\right) \log \left(\frac{m_r^2}{\Lambda^2}\right) \\[0.05cm]
& \hspace{0.6cm} +\left(m_r^2-m_{r'}^2\right) \left(-6 m_r^3 m_{r'}-3 m_r^2 m_{r'}^2+6 m_r m_{r'}^3+m_{r'}^4+4 m_r^4\right) \bigg]\, .
\end{split}
\end{align}
Unlike $C(m_r,m_{r'})$, the functions  $A(m_r,m_{r'})$ and $B(m_r,m_{r'})$ are symmetric in their arguments, as  can be easily verified by inspection.
The effective  vertex $hZ\gamma$ is 
proportional to the antisymmetric combination (see for example Eq.~(\ref{eq:cZgamodel2}))
\begin{equation}
\begin{split}
C(m_r,m_{r'})-C(m_{r'},m_r)  = & \, \frac{1}{8 \pi ^2} \, \frac{1}{\left(m_r^2-m_{r'}^2\right)^2} \, \times \\[0.1cm]
& \times \bigg[ \left(m_r^2-m_{r'}^2\right)  \left(-4 m_r  m_{r'}+m_{r'}^2+m_r^2\right) \\
& \hspace{0.8cm} +2 m_r m_{r'} \left(-m_r m_{r'}+m_{r'}^2+m_r^2\right) \log \left(\frac{m_r^2}{m_{r'}^2}\right) \bigg]\, ,
\end{split}
\end{equation}
which is finite (i.e. cutoff independent) as expected by the argument of Section~\ref{sec:fermloops}.

\section{Spectral functions and $SO(5)$ currents}
\label{sec:spectralfunction}
\setcounter{equation}{0}

For completeness we report here the definition of the spectral function of two currents.
One has:
\begin{equation}
\rho_{\mu\nu}(q) \equiv \sum_n \delta^{(4)}(q-p_n) \langle 0| J_\mu(0) | n\rangle \langle n| J_\nu(0) | 0\rangle \, ,
\end{equation}
where the sum is over a complete set of states. By Lorentz covariance,
\begin{equation}
\rho_{\mu\nu}(q) = \frac{1}{(2\pi)^3} \theta(q^0) \left( q_\mu q_\nu - \eta_{\mu\nu} q^2 \right) \rho(q^2)\, ,
\end{equation}
where $\rho(q^2)$ is the spectral function.

At leading order in the number of fields and derivatives, the expression of the $SO(5)$ conserved currents is (we show for simplicity only terms involving
the NG bosons and the fermions):
\begin{equation}
\label{eq:currents}
\begin{split}
J_\mu^a  & = \sum_r \bar\chi_r \gamma^\mu T^a \chi_r + \dots  \qquad \qquad a = a_L, a_R \\[0.1cm]
J_\mu^{\hat a} & = \frac{f}{\sqrt{2}} \partial_\mu \pi^{\hat a} -  \sum_{r,r'} \left( \zeta_{[r,r']} \,\bar\chi_r \gamma^\mu T^{\hat a} \chi_{r'} + h.c. \right) + \dots 
\end{split}
\end{equation}
%

\section{Two-site vs CCWZ fermionic Lagrangian}
\label{sec:fccwz}
\setcounter{equation}{0}

In this Appendix we briefly discuss the relation between the description of the fermion interactions in the CCWZ approach and the so-called ``two-site'' model
Lagrangian (see for example Ref.~\cite{Contino:2006nn}) where  fermions couple to the Higgs boson only through (non-derivative) Yukawa terms.

In the general case, the  CCWZ Lagrangian of composite fermions is written at leading order in the derivative expansion as
\begin{equation}
\label{eq:lagccwzferm}
{\cal L}=\sum_r \bar\chi_r (i \sl{\nabla}-m_r)\chi_r - \sum_{r,r^\prime} \zeta_{[r,r^\prime]}\, \bar\chi_r \sl{d} \, \chi_r^\prime\, , 
\end{equation}
where the sums run over all possible representations $r, r^\prime$ of the unbroken subgroup ${\cal H}$.
If the composite fermions can be arranged into complete multiplets of the global group ${\cal G}$ (which occurs if ${\cal G}$ is linearly realized at high energy) and
all the parameters $\zeta_{[r,r^\prime]}$ are equal to~1,  it is easy to show that the Lagrangian (\ref{eq:lagccwzferm}) can be rewritten in terms of a ``two-site''  
Lagrangian by means of a field redefinition (see also Ref.~\cite{Grojean:2013qca}).
In this limit, Eq.~(\ref{eq:lagccwzferm}) becomes
\begin{equation}
{\cal L}= \bar\psi\gamma^\mu (iD_\mu + U(\pi)^\dagger i D_\mu U(\pi) )\psi -\sum_r m_r   \overline{(P_r\!\cdot \psi)} (P_r\!\cdot\psi)\, ,
\end{equation}
where $\psi \equiv (\chi_1,\chi_2....)$ denotes the (possibly reducible) representation of ${\cal G}$, and 
$P_r$ is a projector on the  representation $r$ of ${\cal H}$, that is: $ P_r\!\cdot\psi \equiv \chi_r$.   
Also, we have used the fact that $d_\mu + E_\mu = -i U(\pi)^\dagger D_\mu U(\pi)$.
We then perform the field redefinition $\Psi\equiv U \psi$, so that $\Psi$ transforms linearly under ${\cal G}$: $\Psi \to g\, \Psi$.
The Lagrangian can be re-expressed as
\begin{equation}
{\cal L}=\bar\Psi\gamma^\mu i D_\mu\Psi - \sum_r  m_r\overline{(P_r \!\cdot U(\pi)^\dagger \Psi )} (P_r \!\cdot U(\pi)^\dagger \Psi )\, ,
\end{equation}
so  that  the Higgs interactions with fermions  now come entirely from the second term and are of non-derivative type.

As an illustrative example, it is instructive to consider the $SO(5)/SO(4)$ case in which the composite fermions fill  a ${\bf 10}$ of $SO(5)$, where 
$\bf{10=(1,3)+(3,1)+(2,2)}$ under $SU(2)_L \times SU(2)_R$.
The field redefinition in this case reads  $\Psi_{10}=U(\pi)\psi_{10}U(\pi)^\dagger$, where $\psi_{10}=(\chi_{(2,2)},\chi_{(1,3)},\chi_{(3,1)})$  
and both $\Psi_{10}$ and $\psi_{10}$ are conveniently described in $5\times 5$ matrix notation. 
After the field redefinition, the Lagrangian  reads:
\begin{equation}
\label{eq:L10}
\begin{split}
{\cal L}= 
& \, \Tr\!\left[ \bar \Psi_{10} \,i \!\!\not\!\partial\, \Psi_{10} \right] - m_+ \, \Tr\!\left[\bar \Psi_{10} \Psi_{10}\right] \\[0.15cm]
& - \left(m_{\rep[2]{2}}-m_+\right) \,\Tr\!\left[\overline{(P_{(2,2)}\!\cdot U^\dagger \Psi_{10} U)} (P_{(2,2)}\!\cdot U^\dagger \Psi_{10} U )\right] \\[0.15cm]
&- m_{-} \,\Tr \Big[\overline{(P_{(3,1)}\!\cdot U^\dagger \Psi_{10} U)} (P_{(3,1)}\!\cdot U^\dagger \Psi_{10} U ) \\[0.1cm]
& \hspace{1.7cm} -\overline{(P_{(1,3)}\!\cdot U^\dagger \Psi_{10} U)} (P_{(1,3)}\!\cdot U^\dagger \Psi_{10} U ) \Big]\, ,
\end{split}
\end{equation}
where we have defined $m_\pm = (m_{\rep[3]{1}} \pm m_{\rep[1]{3}})/2$.
The action of the projectors $P_{(2,2)}$, $P_{(1,3)}$ and $P_{(3,1)}$ on an element of the algebra $M$ is defined as
\begin{equation}
\label{eq:Paction}
\begin{split}
P_{(2,2)} \!\cdot M & \equiv \sum_{\hat a} T^{\hat a} \,\Tr[ T^{\hat a} M] \, , \\
P_{(3,1)} \!\cdot M & \equiv \sum_{a_L} T^{a_L} \,\Tr[ T^{a_L} M] \, , \\
P_{(1,3)} \!\cdot M & \equiv \sum_{a_R} T^{a_R} \,\Tr[ T^{a_R} M] \, .
\end{split}
\end{equation}
The term in the second line of Eq.~(\ref{eq:L10}) can be more conveniently rewritten in terms of the field $\Phi = U(\pi) \Phi_0$, where $\Phi_0 = (0,0,0,0,1)^T$,
by using identities between $SO(5)$ generators. One has: 
\begin{equation}
\begin{split}
\Tr\!\left[\overline{(P_{(2,2)}\!\cdot U^\dagger \Psi_{10} U)} (P_{(2,2)}\!\cdot U^\dagger \Psi_{10} U )\right] 
& = \sum_{\hat a} \Tr\left[ T^{\hat a} U^\dagger \bar \Psi_{10} U \right] \Tr\left[ T^{\hat a} U^\dagger \Psi_{10} U \right] \\
& = 2 \left( U^\dagger \bar \Psi_{10} \Psi_{10} U \right)_{55} = 2\, \Phi^\dagger \bar\Psi_{10} \Psi_{10} \Phi \, ,
\end{split}
\end{equation}
where in the first equality we have made use of Eq.~(\ref{eq:Paction}). 
The term proportional to  $m_{(3,1)}-m_{(1,3)}$ can be rearranged by using the identities
\begin{equation}
\begin{gathered}
\sum_{a_L}\l(T^{a_L}\r)_{ij}\l(T^{a_L}\r)_{kl}-\sum_{a_R}\l(T^{a_R}\r)_{ij}\l(T^{a_R}\r)_{kl} =-\frac{1}{2}\epsilon^{ijkl5} \\[0.2cm]
\epsilon^{ijkl5}U_{ i' i}U_{ j' j} U_{ k' k} U_{ l' l}=\epsilon^{i'j'k'l' n'} U_{n'5} \,\text{det}(U)=\epsilon^{i'j'k'l' n'} U_{n'5} \, .
\end{gathered}
\end{equation}
One can show that
\begin{equation}
\begin{split}
\Tr\Big[ & \overline{(P_{(3,1)}\!\cdot U^\dagger \Psi_{10} U)} (P_{(3,1)}\!\cdot U^\dagger \Psi_{10} U ) \\[0.1cm]
                 & -\overline{(P_{(1,3)}\!\cdot U^\dagger \Psi_{10} U)} (P_{(1,3)}\!\cdot U^\dagger \Psi_{10} U ) \Big]\,= -\frac{1}{2}\epsilon^{mnrpk} \bar{\Psi}_{10}^{mn}\Psi^{rp}_{10}\Phi^k\, .
\end{split}
\end{equation}
Note that this $P_{LR}$-violating term is invariant under $SO(5)$ but not under $O(5)$, which is expected since $P_{LR}$ is an element of $O(5)$ but not of $SO(5)$.~\footnote{We
define $P_{LR} = \text{diag}(-1,-1,-1,+1,+1)$ so that it is unbroken in the $SO(4)$ vacuum.} 
The lagrangian can thus be written as
\begin{equation}
\begin{split}
{\cal L}= &\, \Tr\!\left[ \bar \Psi_{10} \,i \!\!\not\!\partial\, \Psi_{10} \right] - m_+ \Tr\!\left[\bar \Psi_{10} \Psi_{10}\right]
-2\l(m_{(2,2)}-m_+ \r)\Phi^\dagger \bar\Psi_{10}\Psi_{10} \Phi \\[0.1cm]
& +\frac{m_-}{2}\, \epsilon^{mnrpk} \, \bar{\Psi}_{10}^{mn}\Psi^{rp}_{10}\Phi^k\, .
\end{split}
\end{equation}

\section{Calculation of the fermionic contribution to the decay rate $h\to Z\gamma$ in the mass eigenstate basis}
\label{sec:masseigenstate}
\setcounter{equation}{0}

In the main text we have described the calculation of the contribution of  composite fermions to the $h\to Z\gamma$ decay rate by using the
effective field theory approach. We have thus expanded the loop integrals keeping only the leading terms suppressed by two powers of the
NP scale and neglecting more suppressed contributions. Also, we performed our calculation by neglecting the elementary-composite
mixing terms in the fermionic sector, which explicitly violate the Goldstone symmetry. 
It is however possible, and somehow straightforward, to perform a complete
calculation of the 1-loop contribution of heavy fermions to the decay amplitude of $h\to Z\gamma$ without making  approximations.
In this Appendix we describe such a calculation and show that it reduces to the results presented in the text in the proper limit.

\begin{figure}[t]
\begin{center}
\includegraphics[width=0.3\linewidth]{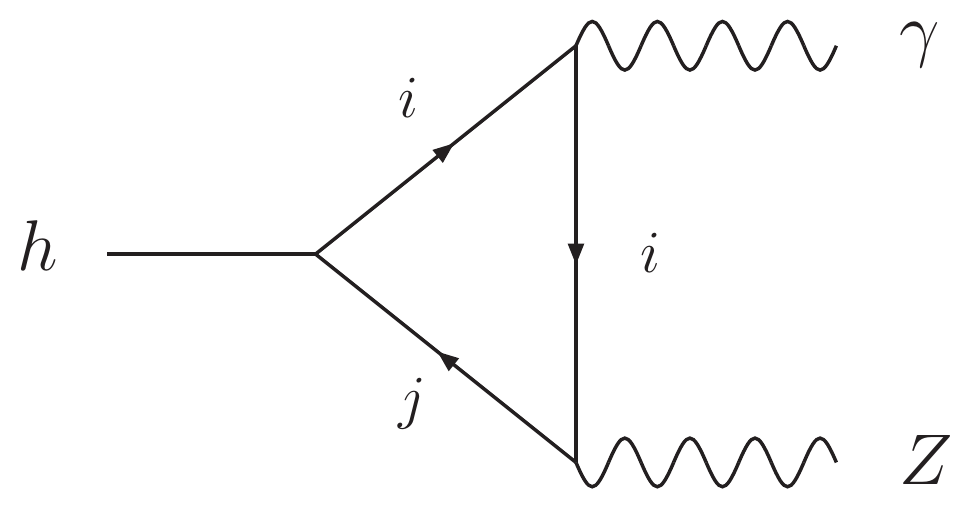}
\caption{\small 
1-loop contribution to $h\to Z\gamma$ from fermion species $i$ and $j$.}
\label{fig:Zgaloop}
\end{center}
\end{figure}

In the SM the fermionic contribution to the decay rate comes from 1-loop diagrams with only one particle species circulating in the loop.
In a generic NP model on the other hand, such as the composite Higgs theories under examination in this paper, there will be several fermions
with the same electromagnetic charge and off-diagonal couplings to the $Z$ and the Higgs boson.
It is thus possible to have two different species of fermions circulating in the same loop for $h\to Z\gamma$, as shown in Fig.~\ref{fig:Zgaloop}.
%
%
In the basis of mass eigenstates and focussing on fermions with the same electric charge, 
the terms of interest in the Lagrangian can be written in full generality as follows
\begin{equation}
\label{eq:generalL}
{\cal L} = \bar\psi^i \left( i\!\!\not\!\partial - m_i\right) \psi^i
+ \frac{1}{2}\bar\psi^i h \l( \lambda^h_{ij} +i\gamma^5 \bar\lambda^h_{ij}  \r) \psi^j
+\frac{1}{2}\bar\psi^i Z_\mu \gamma^\mu \l( \lambda^Z_{ij}+\gamma_5 \bar\lambda^Z_{ij} \r) \psi^j\, , 
\end{equation}
where a sum over all mass eigenstates $i,j$ is left understood and the matrices $\lambda^{h,Z}$,  $\bar\lambda^{h,Z}$ are all hermitian.
Possible derivative interactions of the Higgs with the fermions can always be rewritten as in Eq.~(\ref{eq:generalL}) by integration by parts
and use of the equations of motion. We will show this in detail in the following.
By calculating the diagram of Fig.~\ref{fig:Zgaloop} and summing over $i,j$
one obtains the following decay amplitude
\begin{equation}
\label{nonderferm}
A_{NP}=-\frac{e Q_\psi}{4\pi^2} v \sum_{i,j}
\l[ \lambda^h_{ij}  \lambda^Z_{ji} \, F(m_i,m_j, m_h, m_Z) + i \bar\lambda^h_{ij}  \bar\lambda^Z_{ji} \, F(m_i, -m_j, m_h, m_Z)\r]\, .
\end{equation}
The final result is thus obtained by further summing the contributions from fermions with different electric charge.
The loop function is equal to~\cite{Djouadi:1996yq}
\begin{equation}
\begin{split}
F(m_1,m_2,m_h,m_Z)= \, & \frac{1}{2(m_h^2-m_Z^2)} \,\times \\[0.15cm]
& \times \bigg\{ \frac{m_Z^2 (m_1+m_2)}{(m_h^2-m_Z^2)}\l[B_0(m_Z^2,m_1,m_2)-B_0(m_h^2,m_1,m_2) \r] \\[0.15cm]
&                          \hspace{0.85cm} + \bigg[ -\frac{m_1}{2} (2m_1(m_1+m_2)-m_h^2+m_Z^2 )C_0(m_1,m_1,m_2) \\[0.15cm]
&                          \hspace{1.6cm}                 +(m_1\leftrightarrow m_2)  \bigg]  -(m_1+m_2)
\bigg\}\, ,
\end{split}
\end{equation}
where $B_0(p^2,m_1,m_2),C_0(p_1^2, p_2^2, (p_1+p_2)^2, m_1,m_2,m_3)$ are two- and three-points Passarino-Veltman functions (for a review see Ref.~\cite{Bardin:1999ak}),
and we define for convenience $C_0(m_1,m_2,m_3) \equiv C_0(0, m_Z^2, m_h^2, m_1,m_2,m_3)$.
In the  equal mass limit $m_1=m_2$ the loop function reduces to the SM one (see Eqs.~(\ref{eq:SMamplitudes}), (\ref{eq:SMloopfuncts})):
\begin{equation}
F(m,m,m_h,m_Z) = \frac{1}{2m}(I_1(\tau_f,\lambda_f)-I_2(\tau_f,\lambda_f))\, .
\end{equation}
In the limit of heavy fermions, $m_1^2,m_2^2 \gg m_h^2,m_Z^2$, the loop function reduces to
\begin{equation}
\begin{split}
F(m_1,m_2,0,0)= \, & -\frac{1}{8\left(m_1-m_2\right)^3 \left(m_1+m_2\right)^2}  \, \times \\[0.15cm]
& \times \bigg[  \left(m^2_1-m^2_2\right)  \left(m_1^2-4 m_2 m_1+m_2^2\right) \\[0.15cm]
&  \hspace{1cm} +4 m_1 m_2 \left(m_1^2-m_2 m_1+m_2^2\right) \log \!\left(\frac{m_1}{m_2}\right) \bigg]\, .
\end{split}
\end{equation}

If the Higgs is a NG boson, its interactions to the fermions can be only of derivative type, as shown for example  
in Eq.~(\ref{eq:lagccwzferm}). In the mass-eigenstate basis the Lagrangian can thus be written as
\begin{equation}
\label{eq:NGL}
{\cal L} =\bar\psi^i \left( i\!\!\not\!\partial - m_i\right) \psi^i
+ \frac{1}{2}\bar\psi^i (\partial_\mu h) \gamma^\mu \l( T^h_{ij} + \gamma^5 \bar T^h_{ij}  \r) \psi^j
+\frac{1}{2}\bar\psi^i Z_\mu \gamma^\mu \l( \lambda^Z_{ij}+\gamma_5 \bar\lambda^Z_{ij} \r) \psi^j\, ,
\end{equation}
where $T^h$, $\bar T^h$ are hermitian. By integrating by parts and using the fermions' equations of motion, the above Lagrangian can be re-written as:
\begin{equation}
\begin{split}
{\cal L} = \, & \bar\psi^i \left( i\!\!\not\!\partial - m_i\right) \psi^i
+ \frac{1}{2}\bar\psi^i h \l[ i (m_j-m_i) T^h_{ij} + i \gamma^5 (m_i + m_j) \bar T^h_{ij}  \r] \psi^j \\[0.1cm]
& +\frac{1}{2}\bar\psi^i Z_\mu \gamma^\mu \l( \lambda^Z_{ij}+\gamma_5 \bar\lambda^Z_{ij} \r) \psi^j + O[(h^2\psi^2),(hZ\psi^2)]\, .
\end{split}
\end{equation}
This is of the form (\ref{eq:generalL}) upon identifying 
\begin{equation}
\label{eq:hmatrices}
\lambda^h_{ij} = i (m_j-m_i) T^h_{ij}\, , \qquad \bar\lambda^h_{ij} = (m_i+m_j) \bar T^h_{ij}\, .
\end{equation}
At the 1-loop level the  $O(h^2 \psi^2)$ terms are irrelevant for  $h\to Z\gamma$  and can be safely ignored.
The $O(hZ\psi^2)$ terms also do not contribute to  $h\to Z\gamma$: they  lead to (two-point like) diagrams 
whose loop function has a transverse Lorentz structure,
$(q_\mu q_\nu-g_{\mu\nu} q^2)$,  where $q$ is the photon momentum, hence the corresponding Feynman amplitude vanishes identically for an on-shell photon.
The final expression of the amplitude is thus given by Eq.~(\ref{nonderferm}), with $\lambda^h$ and $\bar\lambda^h$ given by Eq.~(\ref{eq:hmatrices}).

In Section~\ref{sec:fermloops} we have computed the contribution to $h\to Z\gamma$ from pure composite fermions using an effective Lagrangian approach.
For vanishing elementary-composite mixings, the composite  multiplets of $SO(4)$  are mass eigenstates, and Eq.~(\ref{eq:lagccwzferm}) is of the 
form~(\ref{eq:NGL}) with $T^h \propto \zeta \, T^{\hat 4}$ and $\bar T^h =0$. 
The vanishing of $\bar T^h$ follows from our tacit assumption to have the same coupling to the Higgs for both left- and right-handed chiralities of composite
fermions in Eq.~(\ref{eq:lagccwzferm}). By using the above results, in particular Eqs.~(\ref{nonderferm}), (\ref{eq:hmatrices}), and taking the limit of heavy fermion 
masses, the decay amplitude reads
\begin{equation}
A_{NP}=-\frac{eQ_\psi}{4\pi^2} \sum_{i,j} i (m_j-m_i) T^h_{ij} \, \lambda^Z_{ji} \, F(m_i,m_j, 0, 0) \, .
\end{equation}
By summing the contributions from mass eigenstates with different electric charge and using the identity
\begin{equation}
\frac{(m_j-m_{i})}{\pi^2} F(m_i,m_j,0,0)=C(m_i,m_j)-C(m_j,m_i)\, ,
\end{equation}
one finally re-obtains the result of Eq.~(\ref{eq:final1loop}).


\end{document}